\documentstyle[12pt]{article}

\input epsf.sty

\begin{document}
\begin{titlepage}
hep-ph/9702354 \hfill CERN-TH/97-25
\rightline{NTZ 97/6}
\vspace{10mm}

\centerline{\large \bf Scale dependence of the chiral-odd twist-3}
\centerline{\large \bf distributions $h_L(x)$ and $e(x)$}
\vspace{10mm}
\centerline{\bf A.V. Belitsky}

\vspace{5mm}

\centerline{\it Bogoliubov Laboratory of Theoretical Physics}
\centerline{\it Joint Institute for Nuclear Research}
\centerline{\it 141980, Dubna, Russia}

\vspace{10mm}

\centerline{\bf D. M\"uller\footnote{ Permanent address: Institut
f\"ur Theoretische Physik, Universit\"at Leipzig, 04109 Leipzig,
Germany}}

\vspace{5mm}

\centerline{\it Theory Division, CERN,}
\centerline{\it CH-1211 Geneva 23, Switzerland}

\vspace{20mm}

\centerline{\bf Abstract}

\hspace{0.5cm}

We evaluate the complete leading-order evolution kernels for the chiral-odd
twist-3 distributions $e(x)$ and $\widetilde h_L(x)$ of the nucleon. We
establish the connection between the evolution equations in light-cone
position and light-cone fraction representations, which makes a
correspondence between the non-local string operator product expansion and
the QCD-inspired parton model. The compact expression obtained for the local
anomalous dimension matrix coincides with previous calculations. In the
multicolour QCD as well as in the large-$x$ limit the twist-3 distributions
obey simple DGLAP equations. Combining these two limits, we propose improved
DGLAP equations and compare them numerically with the solutions of the exact
evolution equations.

\vfill
\noindent
CERN-TH/97-25 \\
\noindent
February 1997

\end{titlepage}

\section{Introduction}

The factorization theorems \cite{mue89} provide us with powerful tools to
study the processes with large momentum transfers. They give us the
possibility to separate the contributions responsible for physics of large
and small distances. The former are parametrized by the hadron's parton
distribution functions, or by parton correlators in general, which are
uncalculable at the moment from the first principles of the theory, while
the second ones --- hard-scattering sub\-pro\-cess\-es --- can be dealt
perturbatively. The parton distributions are defined in QCD by the target
matrix elements of the light-cone correlators of field operators
\cite{col82}. This representation allows for the estimation of these
quantities by the non-perturbative methods presently available, which are
close to the fundamental QCD Lagrangian \cite{bra95}.

The increasing accuracy of the experimental measurements requires
the unravelling of the twist-3 effects in the hard processes, which
manifest the quantum mechanical interference of partons in the
interacting hadrons. The most important advantage of the twist-3
structure functions is that, while being important for understanding
the long-range quark--gluon dynamics, they contribute at leading
order in $1/Q$ ($Q$ being the momentum of the probe) to certain
asymmetries \cite{jaf90,jaf91} and therefore may be directly
extracted from experiments \cite{abe96}. To confront the theory
with high-precision data, the knowledge of the scale dependence of
the measurable quantities is needed. The $Q^2$ evolution of the parton
distributions \cite{lip72,alt77} can be predicted by exploiting the
powerful methods of renormalization group (RG) and QCD perturbation
theory. The evolution equation of the twist-3 polarized chiral-even
nucleon structure function $g_2(x)$ was extensively discussed in the
literature \cite{lip84}--\cite{mul96} as well as its
solution in the multicolour QCD and asymptotical values of orbital
momentum $n\to\infty$ ($x \to 1$ region) \cite{ali91}.

In this paper we address the study of the evolution of other twist-3
structure functions of the nucleon: chiral-odd distributions $e(x)$ and
$h_L(x)$ \cite{jaf91}, which open a new window to explore the nucleon
content. The present study was induced by several reasons and pursues
various goals. First, while the anomalous dimensions of the local operators
corresponding to the moments of the chiral-odd distributions were computed
for the real QCD case \cite{koi95}, the kernels of the evolution equations
were derived only in the multicolour limit \cite{bbkt96}. From the point of
view of experimental measurements, the knowledge of the evolution of the
whole $x$-dependent distribution is welcome. However, exact equations that
takes into account the ${\cal O}(1/N_c^2)$ effects are lost, although we may
expect sizeable $1/N_c^2$ effects for the small-$x$ behaviour of the
distribution functions, provided the non-leading terms in $N_c$ yield the
rightmost singularity in the complex $n$-plane of the angular momentum with
respect to the leading-order result. Second, up to now the relation between
different formulations of the evolution equations in the light-cone fraction
\cite{lip84,lip85} and light-cone position \cite{bb88,mul94} representations
was obscure. So, the aim of the present paper is two fold: to fill the gap
in our knowledge of the exact (taking into account of the $1/N_c^2$ effects)
higher-twist evolution equations for chiral-odd distributions and to provide
the relation between different calculational approaches. We attempt here to
clarify these issues.

The outline of the paper is the following. We start with the construction of
the basis of the twist-3 chiral-odd polarized and unpolarized correlation
functions, mixed under the re\-nor\-ma\-li\-za\-tion group evolution, and
find the equations that govern their scale dependence in the momentum
fraction as well as in the light-cone position representations in Section 2.
In Section 3 we present the Fourier transformation of the evolution kernels,
which simplifies the transition from one representation to another. Section
4 is devoted to the construction of the generalized DGLAP equation in mixed
representation for the twist-3 distributions, which is the most useful for
solving the RG equations in the large-$N_c$ limit as well as for
asymptotical values of the orbital momentum~$n$. The corresponding
analytical solution, as well as a numerical study of exact equations is
performed in Section 5. The final Section contains a discussion and
concluding remarks.

To make the discussion complete and more transparent, we include several
appendices. In Appendix A we present definitions and some properties of the
$\Theta$-functions that appeared in the formulation of the evolution
equations in the momentum space. In Appendix B the evolution equations for
the redundant basis of operators are found in the Abelian gauge theory. This
shows the self-consistency of the whole approach. The anomalous-dimension
matrix of local twist-3 operators obtained from the evolution equations we
have derived in the main text are written in Appendix C. It coincides with
results known in the literature \cite{koi95}.

\section{Evolution of chiral-odd twist-3 correlation functions}

As we mentioned in the Introduction, the parton distribution
functions in QCD are defined by the Fourier transforms along
the null-plane of the forward matrix element of the parton field
operators product separated by an interval $\lambda$ on the light
cone:
\begin{equation}
\label{two-corr}
{\cal F}(\lambda) \equiv {\cal F}(\lambda, 0)
= \phi^* (0) \phi (\lambda n),
\end{equation}
where $\phi$ denotes a quark $\psi$ or a gluon field $B_\mu$. We
suppress the dependence on the renormalization scale 
$\mu_R$, necessary to make this equation well defined in field
theory. Throughout the paper we use the ghost-free
$B_+ \equiv B_\mu n^\mu = 0$ gauge. Here $n$ is a light-cone
vector $n^2 = 0$ normalized with respect to the four-vector
$P = p + \frac{1}{2}M_h^2n$ of the parent hadron
$h$ of mass $M_h$, i.e. $nP=1$, and $p$ is a null vector
along the opposite tangent to the light cone such that $p^2 = 0$,
$np = 0$. In any covariant gauge a path-ordered link factor should
be inserted between the $\phi$-fields so as to maintain the gauge
invariance of the physical quantities. The Fourier transformations
from the coordinate to the momentum space and vice versa are
given by
\begin{equation}
\label{Fourier-2}
F(x) = \int \frac{d\lambda}{2\pi} e^{i\lambda x}
\langle h |
{\cal F} (\lambda) | h \rangle ,
\hspace{0.5cm}
\langle h | {\cal F} (\lambda) | h \rangle
= \int dx e^{-i\lambda x} F(x).
\end{equation}
Both of these representations display the complementary aspects
of the factorization. The light-cone position representation is
suitable to make contact with the operator product expansion (OPE)
approach, while the light-cone fraction representation is
appropriate for establishing the language of the parton model.
Throughout the paper we will use the light-cone position and
the light-cone fraction representations in parallel.

The multiparton distributions corresponding to the interference
of higher Fock components in the hadron wave functions that emerge
at the twist-3 level are the generalizations of (\ref{two-corr})
to 3-parton fields
\begin{equation}
\label{three-corr}
{\cal F} (\lambda, \mu) \equiv {\cal F} (\lambda, 0, \mu)
= \phi^* (\mu n) \phi (0) \phi (\lambda n).
\end{equation}
We do not display the quantum numbers of the field operators
since they are not of relevance at the moment. The direct and
inverse Fourier transforms are
\begin{equation}
\label{Fourier-3}
F(x, x') = \int \frac{d\lambda}{2\pi}\frac{d\mu}{2\pi}
e^{i\lambda x - i\mu x'}
\langle h | {\cal F} (\lambda , \mu) | h \rangle ,\hspace{0.5cm}
\langle h | {\cal F}(\lambda, \mu) | h \rangle
= \int dxdx' e^{-i\lambda x + i\mu x'} F(x, x').
\end{equation}
The variables $x$ and $x'$ are the momentum fractions of incoming
$\phi$ and outgoing $\phi^*$ partons, respectively. The
restrictions on their physically allowed values come from the
support properties of the multiparton distribution functions
discussed at length in Ref. \cite{jaf83}, namely $F(x, x')$
vanishes unless $0 \leq x \leq 1$, $0 \leq x' \leq 1$.

Beyond the leading-twist level the intuitive parton-like picture is
not so immediate, as one usually starts with an overcomplete set of
correlation functions. However, the point is that the equations of
motion for field operators imply several relations between
correlators, and the problem of construction of the simpler operator
basis is reduced to an appropriate exploitation of these equalities.
The guiding line to disentangle the twist structure is clearly
seen in the light-cone formalism of Kogut and Soper \cite{kog70}.
Consider, for instance, the correlators containing two quarks
$\bar\psi\psi$. Then decomposing the Dirac field into ``good"
and ``bad" components with Hermitian projection operators
${\cal P}_{\pm} = \frac{1}{2}\gamma_{\mp}\gamma_{\pm}$:
$\psi_{\pm} = {\cal P}_{\pm} \psi$, we have three possible
combinations $\psi^\dagger_+\psi_+$, $\psi^\dagger_+\psi_-
\pm \psi^\dagger_-\psi_+$, and $\psi^\dagger_-\psi_-$, which are
of twist 2, 3 and 4, respectively. The origin of this
counting lies in the dynamical dependence of the ``bad" components
of the Dirac fermions
\begin{equation}
\label{bad}
\psi_- = -\frac{i}{2}\partial^{-1}_+
\left(
i\not\!\!D_\perp + m
\right) \gamma_+ \psi_+.
\end{equation}
These components depend on the underlying QCD dynamics, i.e.
they implicitly involve extra partons and thus correspond to
the generalized off-shell partons, which carry the transverse
momentum. For this reason we come back to the on-shell massless
collinear partons of the naive parton model, but supplemented with
multiparton correlations through the constraint (\ref{bad}).
The operators constructed from the ``good" components
only were named quasi-partonic \cite{lip85}. The advantage of
handling them is that they endow the theory with a parton-like
interpretation for higher twists.

At the twist-3 level the nucleon has two chiral-odd distributions
$e$ and $h_L$ \footnote{In the language of OPE the local twist-3
as well as twist-2 operators contribute to the matrix elements
of the distribution function $h_L$. Equation (\ref{tw-2-3}) is a
consequence of this fact.}, which can be measured in the polarized
Drell--Yan and semi--inclusive DIS processes. Under QCD evolution,
they couple to complicated quark--gluon operators and correlation
functions, depending on the quark mass and intrinsic transverse
momentum. We will treat below the unpolarized and polarized cases
separately.

\subsection{Unpolarized distributions}

In the unpolarized case we define the following redundant basis
chiral-odd twist-3 correlation functions:
\begin{eqnarray}
\label{DefUnpoCF-e}
&&\hspace{-0.7cm}e(x)
=\frac{x}{2}\int \frac{d\lambda}{2\pi}
e^{i\lambda x}
\langle h | \bar \psi (0)  \psi (\lambda n)|h \rangle ,
\\
\label{DefUnpoCF-M}
&&\hspace{-0.7cm}M(x)
=\frac{1}{2}\int \frac{d\lambda}{2\pi}
e^{i\lambda x}
\langle h | \bar \psi (0) m\gamma_+  \psi (\lambda n)|h \rangle ,
\\
\label{DefUnpoCF-D1}
&&\hspace{-0.7cm}D_1 (x,x')
=\frac{1}{2}\int \frac{d\lambda}{2\pi} \frac{d\mu}{2\pi}
e^{i\lambda x-i\mu x'}
\langle h | \bar \psi (\mu n)
{\rm g} \gamma_+\! \not\!\! B^\perp(0)
\psi (\lambda n)|h \rangle ,
\\
\label{DefUnpoCF-D2}
&&\hspace{-0.7cm}D_2 (x',x)
=\frac{1}{2}\int \frac{d\lambda}{2\pi} \frac{d\mu}{2\pi}
e^{i\mu x' - i\lambda x}
\langle h | \bar \psi (\lambda n)
{\rm g}\! \not\!\! B^\perp(0) \gamma_+
\psi (\mu n)|h \rangle.
\end{eqnarray}
The functions $D_1$ and $D_2$ are related by complex conjugation
$\left[ D_1 (x,x') \right]^* = D_2 (x',x)$. The quantities
determined by these equations form a closed set under
renormalization; however, they are not independent, since there
is a relation between them due to the equation of motion for the
Heisenberg fermion field operator:
\begin{equation}
\label{EOM-e}
e(x)-M(x)-\int dx' D(x,x')=0,
\end{equation}
where we have introduced the convention
\begin{eqnarray}
&&D (x,x')
=\frac{1}{2}
\left[ D_1 (x,x') + D_2 (x',x) \right].
\end{eqnarray}
This function is real-valued and antisymmetric with respect to the
exchange of its arguments:
\begin{eqnarray}
\left[D(x,x') \right]^* = D(x,x') , \qquad  D (x,x') = - D (x',x).
\end{eqnarray}
In Appendix B we present a set of RG equations for the correlation
functions determined by Eqs. (\ref{DefUnpoCF-e})--(\ref{DefUnpoCF-D2})
derived in Abelian gauge theory. The relation (\ref{EOM-e}) provides
a strong check of our calculations\footnote{This fact follows from
general renormalization properties of gauge-invariant operators
as one expects that the counter term for the equation of motion
operator can be given only by the operator itself. Its matrix
element, being taken with respect to the physical
state, decouples completely from the renormalization group
evolution.}. It allows the reduction of the RG analysis to the study of
scale dependence of the three-parton $D$ and mass-dependent $M$
correlators only.

The function $D(x,x')$ is gauge-variant provided we use the gauge
other than light-cone; therefore, we are forced to introduce the
gauge-invariant quantity
\begin{eqnarray}
\label{GIDefUnpoCF}
Z(x,x') = (x-x')D(x,x').
\end{eqnarray}
Using the advantages of the light-cone gauge, where the gluon field
is expressed in terms of the field strength tensor (the residual
gauge degrees of freedom are fixed by imposing an antisymmetric
boundary conditions on the field, which allows unique inversion):
\begin{equation}
B_{\mu}(\lambda n) =\partial^{-1}_+ G_{+ \mu}(\lambda n)
=\frac{1}{2}\int_{-\infty}^{\infty}dz
\epsilon (\lambda - z)G_{+\mu}(z),\label{G}
\end{equation}
and taking into account the relation
\begin{equation}
\frac{1}{2} \int \frac{d\lambda}{2\pi}
e^{\pm i \lambda x} \epsilon (\lambda - z)
= \pm \frac{i}{2\pi} {\rm PV} \frac{1}{x} e^{\pm i z x},
\end{equation}
we can easily obtain from Eqs.
(\ref{DefUnpoCF-D1}) and (\ref{DefUnpoCF-D2}) the definition
of the gauge-invariant quantities in terms of three-particle
string operators. Generically
\begin{eqnarray}
\label{DefUnpolLRO-Z}
Z(x,x') =
\frac{1}{2}
\int \frac{d\lambda}{2\pi}\frac{d\mu}{2\pi}
e^{i\lambda x - i\mu x'}
\langle h |
{\cal Z}(\lambda,\mu) + {\cal Z}(-\mu,-\lambda)
| h \rangle,
\end{eqnarray}
where
\begin{eqnarray}
\label{DefLRO-Z}
{\cal Z}(\lambda , \mu) \equiv {\cal Z}(\lambda ,0, \mu)
= \frac{1}{2}
\bar \psi (\mu n) {\rm g} G_{+ \rho} (0)
\sigma^\perp_{\rho +} \psi (\lambda n).
\end{eqnarray}
In the same way, for a mass-dependent non-local string operator
\begin{eqnarray}
\label{DefLRO-M}
{\cal M}^j(\lambda) \equiv {\cal M}^j(\lambda, 0)
=\frac{m}{2}\bar \psi (0) \gamma_+ (iD_+ (\lambda) )^j
\psi (\lambda n),
\end{eqnarray}
the Fourier transform is
\begin{eqnarray}
\label{DefUnpolLRO-M}
M^j (x) &=& x^j M (x)
= \int \frac{d\lambda}{2\pi}
e^{i\lambda x}
\langle h |{\cal M}^j( \lambda ) |h \rangle .
\end{eqnarray}

For the spin-dependent scattering discussed below, the only
difference is that one should insert also a $\gamma_5$-matrix
between the fields in the definitions of the string operators
(\ref{DefLRO-Z}), (\ref{DefLRO-M}).

\subsection{Polarized distributions}

Analogously, the set of correlation functions for the polarized
case is as follows:
\begin{eqnarray}
&&\hspace{-0.7cm}h_1(x) \label{DefPoCF-h_1}
=\frac{1}{2}S_\sigma^\perp \int \frac{d\lambda}{2\pi}
e^{i\lambda x}
\langle h | \bar \psi (0)i \sigma^{\ \,\perp}_{+ \sigma} \gamma_5
\psi (\lambda n)|h \rangle ,\\
&&\hspace{-0.7cm}h_L(x) \label{h_L}
=\frac{x}{2}\int \frac{d\lambda}{2\pi}
e^{i\lambda x}
\langle h | \bar \psi (0)i \sigma_{+ -} \gamma_5
\psi (\lambda n)|h \rangle ,\\
&&\hspace{-0.7cm}\widetilde M(x)
=\frac{1}{2}\int \frac{d\lambda}{2\pi}
e^{i\lambda x}
\langle h | \bar \psi (0) m\gamma_+ \gamma_5
\psi (\lambda n)|h \rangle ,\\
&&\hspace{-0.7cm}K(x) \label{K}
=\frac{1}{2}\int \frac{d\lambda}{2\pi}
e^{i\lambda x}
\langle h | \bar \psi (0) i\gamma_+\!\not\!\partial_\perp
\gamma_5
\psi (\lambda n)|h \rangle ,\\
&&\hspace{-0.7cm}\widetilde D_1 (x,x')
=\frac{1}{2}\int \frac{d\lambda}{2\pi} \frac{d\mu}{2\pi}
e^{i\lambda x-i\mu x'}
\langle h | \bar \psi (\mu n)
{\rm g} \gamma_+\! \not\!\! B^\perp(0) \gamma_5
\psi (\lambda n)|h \rangle ,\\
&&\hspace{-0.7cm\widetilde }D_2 (x',x) \label{DefPoCF-D2}
=\frac{1}{2}\int \frac{d\lambda}{2\pi} \frac{d\mu}{2\pi}
e^{i\mu x' - i\lambda x}
\langle h | \bar \psi (\lambda n)
{\rm g} \gamma_+\! \not\!\! B^\perp(0) \gamma_5
\psi (\mu n)|h \rangle,
\end{eqnarray}
where $S_\sigma^\perp$ denotes the transverse polarization vector
of the hadron $h$ ($S^2=-M_h^2$). The derivative in the correlation
function $K(x)$ acts on the quark field before setting its argument
on the light cone.

Besides the identity arising from the equation of motion
\begin{equation}
\label{EOM-h_L}
h_L(x)-\widetilde M(x) - K(x)
-\int dx'\widetilde D(x,x')=0,
\end{equation}
there is an equation provided by Lorentz invariance
\begin{equation}
\label{tw-2-3}
2xh_1 (x)
=2h_L (x) - x \frac{\partial}{\partial x}K(x)
-2x\int dx' \frac{\widetilde D(x,x')}{(x' - x)}.
\end{equation}
It means that both parts of this equality are expressed in terms
of matrix elements of different components of one and the same
twist-2 tensor operator. Again we have introduced the $C$-even
quantity $\widetilde D$, which has the properties
\begin{equation}
\left[ \widetilde D (x,x') \right]^* = \widetilde D (x, x'),
\quad
\widetilde D (x,x') = \widetilde D (x',x).
\end{equation}
Combining the two Eqs.\ (\ref{EOM-h_L}) and (\ref{tw-2-3}) we
can obtain the following relation between the correlators:
\begin{equation}
\left( 2 - x \frac{\partial}{\partial x} \right) h_L(x)
=2x h_1(x) - x \frac{\partial}{\partial x} \widetilde M (x)
+ \int dx' \frac{x}{x' - x}
\left[ \frac{\partial}{\partial x}
- \frac{\partial}{\partial x'} \right]
\widetilde Z (x, x'),
\end{equation}
where $\widetilde Z (x, x')$ is a gauge-invariant quantity
introduced by Eq. (\ref{GIDefUnpoCF}), but for the function
$\widetilde D (x, x')$. Solving the differential equation
with respect to $h_L(x)$ the integration constant can be
found from the support properties of the distribution:
$h_L(x) = 0$ for $|x|\geq 1$. The solution is
\begin{equation}
\label{sumrule}
h_L(x)
= 2x^2 \int_{x}^{1}\frac{d\beta}{\beta^2} h_1(\beta)
+ \widetilde M (x)
- 2x^2\int_{x}^{1}\frac{d\beta}{\beta^3} \widetilde M (\beta)
+ x^2 \int_{x}^{1} \frac{d\beta}{\beta^2}
\int \frac{d\beta'}{\beta' - \beta}
\left[ \frac{\partial}{\partial \beta}
- \frac{\partial}{\partial \beta'} \right]
\widetilde Z (\beta, \beta').
\end{equation}
A similar relation was found by Jaffe and
Ji \footnote{Corresponding expressions in Ref. \cite{jaf91}
contain misprints.} in Ref. \cite{jaf91}. Here
the dynamical twist-3 contribution is explicitly related to
the particular integral of the three-parton correlation function
$\widetilde Z$. In terms of local operators it looks like
\begin{equation}
(n + 3)[h_L]_n = 2 [h_1]_{n+1} + (n + 1) \widetilde M_n
+\sum_{l=1}^{n} (n-l+1) \widetilde Z_n^l ,
\end{equation}
and the definition of moments of distribution functions is given
by Eq. (\ref{mom-2-3}).

As before, excluding the functions (\ref{h_L}) and (\ref{K}), and using
the relations (\ref{EOM-h_L}) and (\ref{tw-2-3}), we can chose the basis
of independent functions in the form: $h_1(x)$, $\widetilde M(x)$,
$\widetilde D(x, x')$.

\subsection{Evolution equations}

Note that in the leading logarithmic approximation the evolution
equations that govern the $Q^2$-dependence of the three-particle
correlation functions are the same, discarding the mixing with the
quark mass operator. Therefore we omit the ``tilde" sign in what
follows. We evaluate the evolution equations using the different
approaches described in \cite{lip84,lip85} and \cite{mul96} and
keep to general remarks.

\begin{figure}[htb]
\mbox{
\hspace{1cm}
\epsffile{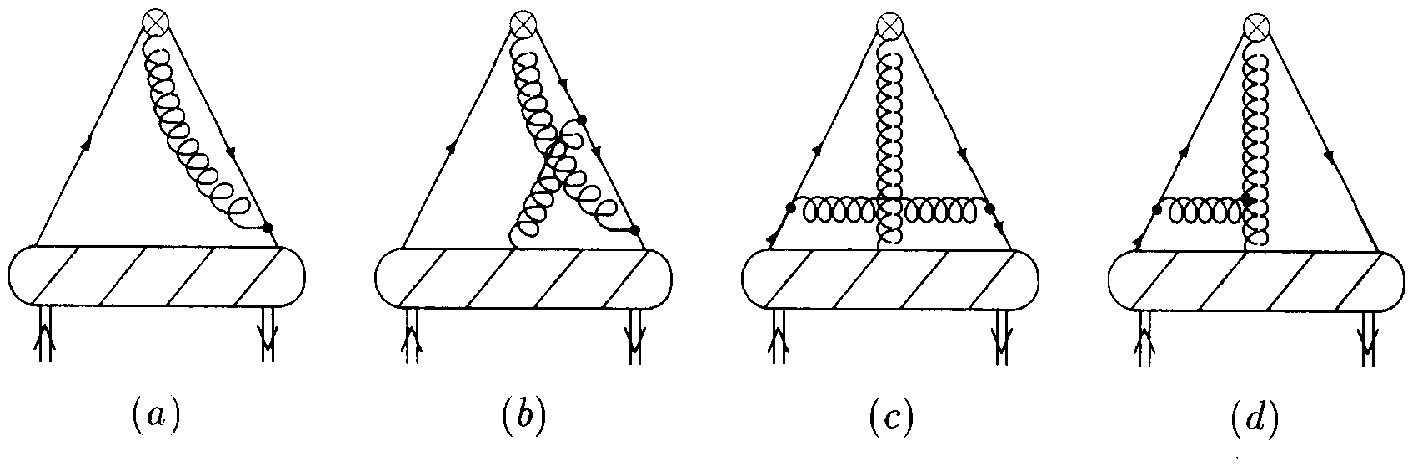}}
\vspace{0.1cm}
{\caption{\label{three-pa}
The one-loop renormalization of the three-parton correlation
functions.
}}
\end{figure}

The peculiar feature of the light-like gauge is the presence of
the spurious IR pole $1/k_+$ in the density matrix of the gluon
propagator:
\begin{eqnarray}
D_{\mu\nu}(k)=\frac{d_{\mu\nu}(k)}{k^2+i0}, \quad
d_{\mu\nu}= g_{\mu\nu}-\frac{k_\mu n_\nu + k_\nu n_\mu}{k_+}.
\end{eqnarray}
The central question is how to handle this unphysical pole
when $k_+ = 0$. In the calculation of the relevant Feynman
diagrams shown in Fig.~\ref{three-pa} (and trivial self-energy
insertions into external legs) we assume two different approaches,
which employ the principal value (PV) and Mandelstam--Leibbrandt
prescriptions (ML) \cite{lei87}:
\begin{eqnarray}
{\rm PV}\frac{1}{k_+} &=& \frac{1}{2}
\left\{ \frac{1}{(kn) + i0} + \frac{1}{(kn) - i0} \right\}, \\
{\rm ML}\frac{1}{k_+} &=& \frac{(kn^*)}{(kn)(kn^*) + i0},
\end{eqnarray}
with the arbitrary four-vector $n^*$ satisfying $n^{*2} = 0$,
$nn^* = 1$ (without loss of generality, we can put
it equal to $p$). The first prescription will be used in the
momentum space \cite{lip84,lip85,bel96}, the second one in the
coordinate space formulation \cite{mul96}. As a by-product we verify
that both of them do lead to the same result.

In the light-cone fraction representation we get for the
correlation function $D(x,x')$:
\begin{eqnarray}
\label{eveqz}
&\dot D&\hspace{-0.3cm}(x , x')=
-\frac{\alpha}{2\pi}
\Biggl\{
-C_F \frac{(x - x')}{x x'}
\left[
x' M (x) \Theta_{11}^0 (x', x' - x)
\pm
x M (x') \Theta_{11}^0 (x, x - x')
\right]
\nonumber\\
&&+
\int d\beta
\Biggl(
C_F  D (\beta , x')\frac{x}{x'}
\Theta_{11}^0 (x, x - x')
+\frac{C_A}{2}
\biggl(
[D (\beta , x') - D (x , x')]\frac{x}{(x - \beta)}
\Theta_{11}^0 (x, x - \beta) \nonumber\\
&&+ [D (\beta + x', x') - D (x , x')]
\frac{(x - x')}{(x - x' - \beta)}
\Theta_{11}^0 (x - x', x - x' - \beta) \nonumber\\
&&+\frac{(\beta + x - x')}{x'}
\biggl(
D (\beta , x')\frac{x}{(x' - x)}
\Theta_{11}^0 (x , x - \beta)
+
D (\beta + x', x')\Theta_{11}^0
(x - x' , x - x' - \beta)
\biggr)
\biggr)\nonumber\\
&&+
\left(
C_F - \frac{C_A}{2}
\right)
\biggl(
D(\beta , x')\frac{(\beta + x - x')}{(x' - x)}
\Theta_{111}^0 (x, x - x' , x - x' + \beta)\nonumber\\
&&+
[D (\beta , x' - x + \beta) - D (x, x')]\frac{x}{x - \beta}
\Theta_{11}^0 (x , x - \beta)
\biggr)
\Biggr)\nonumber\\
&&+
\int d\beta'
\Biggl(
C_F  D (x , \beta')\frac{x'}{x}
\Theta_{11}^0 (x', x' - x)
+\frac{C_A}{2}
\biggl(
[D (x , \beta') - D (x , x')]\frac{x'}{(x' - \beta')}
\Theta_{11}^0 (x', x' - \beta') \nonumber\\
&&+ [D (x , \beta' + x) - D (x , x')]
\frac{(x' - x)}{(x' - x - \beta')}
\Theta_{11}^0 (x' - x, x' - x - \beta') \nonumber\\
&&+\frac{(\beta' + x' - x)}{x}
\biggl(
D (x , \beta')\frac{x'}{(x - x')}
\Theta_{11}^0 (x' , x' - \beta')
+
D (x , \beta' + x)\Theta_{11}^0
(x' - x , x' - x - \beta')
\biggr)
\biggr)\nonumber\\
&&+
\left(
C_F - \frac{C_A}{2}
\right)
\biggl(
D( x ,\beta' )\frac{(\beta' + x' - x)}{(x' - x)}
\Theta_{111}^0 (x', x' - x , x' - x + \beta')\nonumber\\
&&+
[D ( x - x' + \beta' , \beta') - D (x, x')]\frac{x'}{x' - \beta'}
\Theta_{11}^0 (x' , x' - \beta')
\biggr)
\Biggr) - \frac{3}{2}C_F D(x,x')
\Biggr\},
\end{eqnarray}
and for the mass-dependent correlation function we have
\begin{equation}
\label{mass}
\dot M(x)
= - C_F \frac{\alpha}{2\pi}\int d \beta M(\beta)
\left\{
2
\left[
\frac{ \beta }{( x - \beta )}\Theta_{11}^0 (x, x - \beta)
\right]_+
+
\frac{\beta + x}{\beta} \Theta_{11}^0 (x, x-\beta)
\right\},
\end{equation}
where we have used the dot as short-hand for the logarithmic
derivative with respect to the renormalization scale
$\hspace{0.1cm}\dot{}=\mu_R^2\,{\partial}/{\partial\mu_R^2}$
and the standard plus-prescription fulfilling $\int dx [...]_+
= 0$ [for a definition see Eq. (\ref{plus})]. The
kernel of the last equation resembles the non-singlet splitting
function of (un)po\-la\-rized scattering up to the additional
term $\frac{\alpha}{2\pi} \frac{3}{2}C_F\delta(x - \beta)$, which
is the one-loop renormalization constant of the quark mass taken
with a minus sign. An explicit form of the $\Theta$-functions is
given in Appendix A. Throughout the paper the plus and minus
signs in the mass-operator term correspond to the functions $D$
(for $e$) and $\widetilde D$ (for $h_L$), respectively.

For the string operators (or their matrix elements) we obtain
the following compact RG equation:
\begin{eqnarray}
\label{evolequaZ}
\dot{\cal Z}(\lambda , \mu) &=&
\frac{\alpha}{2 \pi}
\int_{0}^{1} dy \int_{0}^{\bar y} dz
\biggl\{
C_F {\bar y}^2 \delta (z)
\left[
{\cal M}^1 (\lambda - \mu y)
\pm
{\cal M}^1 (\lambda y - \mu)
\right] \nonumber \\
&+&\frac{C_A}{2}
\left[  2\bar z + [N(y,z)]_+ -\frac{7}{4} \delta(\bar y) \delta(z)
\right]
\left[ {\cal Z}(\lambda y, \mu - \lambda z)
+ {\cal Z}(\lambda - \mu z, \mu y) \right] \nonumber\\
&+& \left( C_F - \frac{C_A}{2} \right)
\biggl[
\left[ [L(y,z)]_+ - \frac{1}{2} \delta(y) \delta(z) \right]
{\cal Z}(\lambda \bar z + \mu z , \mu \bar y + \lambda y)
\nonumber\\
&-& 2z \left[
{\cal Z}(-\lambda y , \mu - \lambda \bar z)
+ {\cal Z}(\lambda - \mu \bar z , -\mu y)
\right]
\biggr]
\biggr\},
\end{eqnarray}
with
\begin{eqnarray}
&&[N(y,z)]_+ = N(y,z)
- \delta(\bar y) \delta(z)
\int_{0}^{1} dy' \int_{0}^{\bar y'} dz' N(y',z'), \hspace{0.5cm}
N(y,z) = \delta (\bar y-z) \frac{y^2}{\bar y}
+ \delta (z) \frac{y}{\bar y},\nonumber\\
&&[L(y,z)]_+ = L(y,z)
- \delta(y) \delta(z)
\int_{0}^{1} dy' \int_{0}^{\bar y'} dz' L(y',z'), \hspace{0.5cm}
L(y,z) = \delta (y) \frac{\bar z}{z} + \delta (z) \frac{\bar y}{y}.
\end{eqnarray}
The equations written so far should be supplemented by the following
\begin{eqnarray}
\label{mass-NL}
\dot{\cal M}^1 (\lambda) &=& \frac{\alpha}{2\pi}C_F
\int_{0}^{1}dy
\left\{ \left[ \frac{2}{\bar y} \right]_+ - 2 - y - y^2 \right\}
{\cal M}^1 (\lambda y) ,\\
\dot h_1 (\lambda) &=& \frac{\alpha}{2\pi}C_F
\int_{0}^{1}dy
\left\{ \left[ \frac{2}{\bar y} \right]_+
- 2 + \frac{3}{2}\delta (\bar y) \right\}
h_1 (\lambda y).
\end{eqnarray}
The last one, when transformed to the momentum space using the
formulae of the next section, coincides with the result obtained in
Ref. \cite{art90}.

\section{From light-cone position to light-cone fraction \\
re\-pre\-sen\-ta\-tions}

Having at hand the evolution equations in different representations
for the same quantities, it is instructive to relate the kernels in
both cases. Such a bridge can be easily established using the
Fourier transformation for the parton distribution functions given
by Eqs. (\ref{Fourier-2}) and (\ref{Fourier-3}).

First, we come to the simpler case of the two-particle correlation
functions ${\cal F}$. The evolution equation in
the light-cone position space is of the following generic form
\begin{equation}
\dot{\cal F} (\lambda)
= \int_{0}^{1} dy {\cal K} (y) {\cal F} (\lambda y),
\end{equation}
where ${\cal K} (y)$ is an evolution kernel in the coordinate space.
By exploiting the definitions (\ref{Fourier-2}) we can recast the
Fourier transform on the language of two-particle evolution kernels.
In this way we find the direct transformation
\begin{equation}
K(x,\beta) = \int_{0}^{1} dy {\cal K}(y) \delta (x - y\beta).
\end{equation}
And using the general formula
\begin{equation}
\label{theta}
\int_{0}^{1} dy f(y) \delta (x - y\beta)
= f\left( \frac{x}{\beta} \right)
\Theta^0_{11} (x, x - \beta).
\end{equation}
with the familiar $\Theta$-function of the momentum space formulation,
we can observe that the RG equations for two-parton correlators
derived in the previous section indeed coincide. The inverse
transformation can be done
\begin{equation}
\int \frac{dxd\beta}{2\pi} e^{-i\lambda x + i\mu \beta}
K(x, \beta) = \int_{0}^{1}dy {\cal K}(y) \delta (\mu - y \lambda)
\end{equation}
with the help of the formula
\begin{equation}
\int \frac{dxd\beta}{2\pi} f\left( \frac{x}{\beta} \right)
\Theta^0_{11} (x, x - \beta) e^{-i\lambda x + i\mu \beta}
= \int_{0}^{1} dy f(y) \delta (\mu - y \lambda).
\end{equation}

The corresponding transformation for three-particle correlators is a
little bit more involved. The general form of the evolution equation
for the light-cone string operator ${\cal Z}(\lambda,\mu)$ reads
\begin{equation}
\dot{\cal Z}(\lambda , \mu)
=\int_{0}^{1}dy\int_{0}^{\bar y}dz {\cal K}(y,z)
{\cal Z} (\eta_{11}\lambda + \eta_{12}\mu,
\eta_{21}\lambda + \eta_{22}\mu),
\end{equation}
where $\eta_{ij}$ are linear functions of the variables $y$, $z$.
In the momentum fraction representation the evolution equation
looks like
\begin{equation}
\dot Z(x,x') = \int d\beta d\beta'
K (x,x',\beta ,\beta') Z(\beta,\beta').
\end{equation}
Specifying the particular form of the functions $\eta_{ij}$,
we list below the corresponding conversion formulae.

For the $C_A/2$ part of the evolution equation, the Fourier
transformation gives
\begin{equation}
K(x,x',\beta,\beta') = \delta (\beta' - x')
\int_{0}^{1}dy\int_{0}^{\bar y}dz
{\cal K}(y,z)\delta (x - x'z - \beta y).
\end{equation}
The particular contributions are
\begin{eqnarray}
&&\hspace{-0.5cm}{\cal K}_1 (y,z) = \delta (\bar y) \delta (z)
\ \stackrel{FT}{\rightarrow} \ K_1(x,x',\beta,\beta')
= \delta (\beta - x)
\delta (\beta' - x'), \\
&&\hspace{-0.5cm}{\cal K}_2 (y,z) = \delta (z) \frac{y}{\bar y}
+ \delta (\bar y -z)\frac{y^2}{\bar y}
\ \stackrel{FT}{\rightarrow} \ K_2(x,x',\beta,\beta')
= \delta (\beta' - x')
\biggl\{
\frac{x}{\beta - x} \Theta^0_{11} (x, x - \beta) \nonumber\\
&&\hspace{7.8cm}+ \frac{(x - x')^2}{(\beta - x)(\beta - x')}
\Theta^0_{11} (x - x', x - \beta)
\biggr\}, \\
&&\hspace{-0.5cm}{\cal K}_3 (y,z) = 1
\ \stackrel{FT}{\rightarrow} \ K_3(x,x',\beta,\beta') =
\delta (\beta' - x') \Xi_1 (x, x - x', x - \beta) \nonumber\\
&&\hspace{7.8cm}= - \delta (\beta' - x') \Theta^0_{111}
(x, x - x', x - \beta), \\
&&\hspace{-0.5cm}{\cal K}_4 (y,z) = z
\ \stackrel{FT}{\rightarrow} \ K_4(x,x',\beta,\beta') =
\delta (\beta' - x')
\biggl\{
\frac{x - \beta}{x'} \Xi_1 (x, x - x', x - \beta)\nonumber\\
&&\hspace{7.8cm}+ \frac{\beta}{x'} \Xi_2 (x, x - x', x - \beta)
\biggr\}.
\end{eqnarray}
Here we have used (\ref{theta}) and the following useful formula
\begin{eqnarray}
\label{Xi}
&&\Xi_n (x, x-x', x-\beta) \equiv
\int_{0}^{1} dy y^n \Theta^0_{11}
((x-\beta)+y\beta,(x-\beta)-y(x'-\beta))\nonumber\\
&&=\frac{1}{n}\left[ 1
- \left( \frac{\beta - x}{\beta - x'} \right)^n \right]
\Theta^0_{11} (x, x-x')
+\frac{1}{n}\frac{\beta}{x'}
\left[ \left( \frac{\beta - x}{\beta - x'} \right)^n
- \left( \frac{\beta - x}{\beta} \right)^n
\right]
\Theta^0_{11} (x, x-\beta).
\end{eqnarray}

The analogous results for the $(C_F-C_A/2)$ part of the kernel read
\begin{equation}
K(x,x',\beta,\beta') = \delta (\beta' - x')
\int_{0}^{1}dy\int_{0}^{\bar y}dz
{\cal K}(y,z)\delta (x - x'\bar z + \beta y),
\end{equation}
and in particular
\begin{eqnarray}
&&\hspace{-0.5cm}{\cal K}_1 (y,z) = z
\ \stackrel{FT}{\rightarrow} \ K_1(x,x',\beta,\beta') =
\delta (\beta' - x')
\biggl\{
\frac{x' - x - \beta}{x'} \Xi_1 (x' - x, - x, x' - x - \beta)\nonumber\\
&&\hspace{7.8cm}+ \frac{\beta}{x'} \Xi_2 (x' - x, - x, x' - x - \beta)
\biggr\}.
\end{eqnarray}
In addition, we have
\begin{equation}
K(x,x',\beta,\beta') = \int_{0}^{1}dy\int_{0}^{\bar y}dz
{\cal K}(y,z)\delta (x - \beta \bar z + \beta' y)
\delta (x' - \beta'\bar y + \beta z)
\end{equation}
and
\begin{eqnarray}
&&\hspace{-0.5cm}{\cal K}_1 (y,z) = \delta (z) \frac{\bar y}{y}
\ \stackrel{FT}{\rightarrow} \ K_1(x,x',\beta,\beta') =
\delta (\beta - (\beta' - x' + x))
\frac{x'}{\beta' - x'} \Theta^0_{11} (x', x' - \beta').
\end{eqnarray}
These formulae complete the list of transformations. Collecting
particular contributions, we can easily verify that the evolution
equations given by Eqs. (\ref{eveqz}) and (\ref{evolequaZ})
agree with each other. It should be noted that it is sufficient to
have at hand Eqs. (\ref{theta}) and (\ref{Xi}) to perform the
conversion from one representation to another.

\section{Generalized DGLAP-type equations}

To study the large-$N_c$ limit and establish the relation to the
evolution equation for the partially Mellin-transformed operators
introduced in Refs.\ \cite{bb88,ali91,bbkt96}, we proceed to
the generalized DGLAP representation of the evolution equation
for three-particles distributions first given in the second paper
of Ref.\ \cite{mul96}. For this purpose we define a new function,
Fourier-transformed with respect to the $\lambda$ variable only:
\begin{eqnarray}
\label{def3part}
{\scriptstyle\cal Z}(x,u)
= \frac{1}{2} \int \frac{d\lambda}{2\pi}
e^{i\lambda x }
\langle h | {\cal Z}(\bar u \lambda, -u \lambda)
\pm ( u \to \bar u ) |h \rangle ,
\end{eqnarray}
which is even under charge conjugation and depends on the variables
$x$ and $u$. The latter has the meaning of the relative position of
the gluon field on the light cone. For $0\le u \le 1$ the gluon
field lies between the two quark fields. Because of the support
property $|x|\leq {\rm max}(1,|2u-1|)$, the variable $x$ is then
restricted to $|x|\leq 1$ and can be interpreted as an effective
momentum fraction.

The evolution equation for ${\scriptstyle\cal Z}(x,u)$
can be derived in a straightforward way from the RG equation
(\ref{evolequaZ}) for the non-local string operator ${\cal Z}$.
It can be presented in the form of a generalized DGLAP-type
equation:
\begin{eqnarray}
\label{EvoEqux&u}
\dot {\scriptstyle\cal Z}\left( x,u \right)
&=&\frac{\alpha_s}{2\pi}
\int \frac{dy}{y} \int dv
\Bigg\{
P_{{\scriptscriptstyle\cal Z Z}}(y,u,v)
{\scriptstyle\cal Z} \left( \frac{x}{y},v \right)
+
P_{{\scriptscriptstyle\cal Z} m }(y,u,v)
m \left( \frac{x}{v} \right)
\Bigg\}, \\
\label{EvoEqux&u-mass}
\dot m(x)
&=&\frac{\alpha_s}{2\pi}
\int \frac{dy}{y} P_{mm}(y)
m \left( \frac{x}{y} \right).
\end{eqnarray}
Here, $m(x)=xM(x)$ and the integration region is determined both
by the support of ${\scriptstyle\cal Z}(x,u)$ and by the kernels
\begin{eqnarray}
&&P_{\scriptscriptstyle\cal Z Z}(x,u,v) \nonumber\\
&&=\left( C_F - \frac{C_A}{2} \right)
\Bigg[ {\mit\Theta}_1(x,u,v) [L(x,u,v)]_+
-{\mit\Theta}_2(x,u,v) M(x,u,v)
-{1\over 4} \delta(u-v) \delta(\bar x)\Bigg] \nonumber\\
&&+\frac{C_A}{2}{\mit\Theta}_3(x,u,v)
\left[ M(x,u,v) + [N(x,u,v)]_+
-\frac{7}{4} \delta(u-v)\delta(\bar x)\right]
+ {u \rightarrow \bar u \choose v \rightarrow \bar v},\\
&&\nonumber\\
&&P_{{\scriptscriptstyle\cal Z} m}(x,u,v)
=C_F {\bar x}^2\theta(x)\theta(\bar x) \frac{x}{v}
\left[ \delta(v-\bar u - x u) \pm \delta(v-u-x\bar u ) \right],\\
&&\nonumber\\
\label{Pmm}
&&P_{mm}(x)
=C_F \left[
\left[\frac{2}{\bar x}\right]_+ - 2 - x - x^2
\right],
\end{eqnarray}
where the auxiliary functions are defined by:
\begin{eqnarray}
{\mit\Theta}_1(x,u,v)&=&
\theta(x) \theta(u - x v) \theta(\bar u - x \bar v ), \nonumber\\
{\mit\Theta}_2(x,u,v)&=&
\theta\left( -\frac{x\bar v}{\bar u} \right)
\theta\left( \frac{1- x v }{\bar u} \right)
\theta\left( \frac{x - u}{\bar u} \right), \nonumber\\
{\mit\Theta}_3(x,u,v)&=&
\theta\left( \frac{\bar x}{\bar u} \right)
\theta\left( \frac{x \bar v}{\bar u} \right)
\theta\left( \frac{x v - u}{\bar u} \right),
\nonumber\\
L(x,u,v) &=& \frac{u^2}{v (v-u)}\delta(u - x v) ,
\nonumber\\
M(x,u,v) &=& \frac{2 x (1 - x v)}{{\bar u}^3},
\nonumber\\
N(x,u,v) &=& \frac{\bar v \epsilon (\bar u)}{\bar u (v-u)}
\left[ \frac{\bar v}{\bar u} \delta(\bar x)
+ \frac{u^2}{v} \delta(u - x v) \right].
\end{eqnarray}
The plus-prescription for the arbitrary function $A$ is defined by
the equation
\begin{eqnarray}
&&{\mit\Theta}_i(x,u,v) [A(x,u,v)]_+ \nonumber\\
&&= {\mit\Theta}_i(x,u,v) A(x,u,v)- \delta(\bar x) \delta(u-v)
\int_0^1 dx'\int dv'\, {\mit\Theta}_i(x',u,v') A(x',u,v').
\end{eqnarray}
Note that, due to the evolution, the variable $u$ is no longer
restricted to the region $0\le u \le 1$.

Going further, we introduce the Mellin transforms
\begin{equation}
{\scriptstyle\cal Z}^n(u) = \int dx x^{n-1} {\scriptstyle\cal Z}(x,u)
\hspace{0.5cm}\mbox{and}\hspace{0.5cm}
m^n = \int dx x^{n-1} m(x) = \int dx\, x^{n} M(x),
\end{equation}
where $n$ is the complex angular momentum. Operators with different $n$
do not mix with each other and satisfy the evolution equations
\begin{eqnarray}
\label{Mellin-Z}
\dot{\scriptstyle\cal Z}^n(u) &=& \frac{\alpha_s}{2\pi}
\int dv
\left\{
P^n_{{\scriptscriptstyle\cal Z Z}} (u,v)
{\scriptstyle\cal Z}^n (v)
+\delta(u - v)
\left[
P^n_{{\scriptscriptstyle\cal Z} m}(v)
\pm P^n_{{\scriptscriptstyle\cal Z} m} (\bar v)
\right] m^n
\right\}, \\
\dot m^n &=& \frac{\alpha_s}{2\pi}
P_{mm}^n m^n .
\end{eqnarray}
The kernels are given by
\begin{eqnarray}
\label{DefPZZ}
&&P^n_{{\scriptscriptstyle\cal Z Z}}(u,v)
= \left( C_F - \frac{C_A}{2} \right)
\Bigg[ {\mit\Theta}_1(u,v) [L^n(u,v)]_+
- {\mit\Theta}_2 (u,v) M^n_1(u,v)
- \frac{1}{4} \delta(u-v)\Bigg] \nonumber\\
&&+ \frac{C_A}{2}{\mit\Theta}_3(u,v)
\left[ M^n_2(u,v) + [N^n(u,v)]_+ - \frac{7}{4} \delta(u-v) \right]
+ {u \rightarrow \bar{u} \choose v\rightarrow \bar{v}}, \nonumber\\
&& \nonumber\\
&&P^n_{{\scriptscriptstyle\cal Z}m}(v)
= C_F \frac{2 - {\bar v}^{n}[2 + n(2 + (n+1) v)v]}
{n(n+1)(n+2)v^3 }, \nonumber\\
&& \nonumber\\
&&P^n_{mm}
= - C_F \left( S_{n} + S_{n+2} \right),
\end{eqnarray}
where the auxiliary functions read
\begin{equation}
{\mit\Theta}_1(u,v) = \theta(v-u), \quad
{\mit\Theta}_2(u,v)= \theta(- \bar v) \theta(1 - v u), \quad
{\mit\Theta}_3(u,v)= \theta(\bar v)\theta(v-u).
\end{equation}
\begin{eqnarray}
L^n(u,v)
&=& \frac{\epsilon (v)}{v-u}
\left( \frac{u}{v} \right)^{n+1}, \nonumber\\
M_1^n(u,v)
&=& \frac{2}{{\bar u}^3}
\left\{
\frac{1}{n+1} \left[ \frac{1}{v^{n+1}} - u^{n+1} \right]
- \frac{v}{n+2} \left[ \frac{1}{v^{n+2}} -u^{n+2} \right]
\right\},\nonumber\\
M_2^n(u,v)
&=& \frac{2}{{\bar u}^3}
\left\{
\frac{1}{n+1} \left[ 1 - \left(\frac{u}{v} \right)^{n+1} \right]
- \frac{v}{n+2} \left[ 1 - \left(\frac{u}{v} \right)^{n+2} \right]
\right\}, \nonumber\\
N^n(u,v)
&=& \frac{\bar v \epsilon(\bar u)}{\bar u (v-u)}
\left\{ \frac{\bar v}{\bar u}
+ \epsilon (v) \left( \frac{u}{v} \right)^{n+1} \right\}.
\end{eqnarray}
The plus-prescription is defined as
\begin{equation}
{\mit\Theta}_i(u,v) [A^n(u,v)]_+ =
{\mit\Theta}_i (u,v)A^n(u,v)
-\delta(u-v) \int dv' {\mit\Theta}_i(u,v') A^n(u,v').
\end{equation}
It is not difficult to observe that, in multicolour limit, Eq.\ 
(\ref{Mellin-Z}) is exactly reduced to the equation of
Ref. \cite{bbkt96}, which was the starting point of their
analysis. However, we will start from Eq. (\ref{EvoEqux&u})
in the mixed representation and show in the next section that
in the multicolour limit the generalized splitting functions can
be diagonalized.

\section{Solution of the evolution equations in multicolour QCD}

This section is devoted to the solution of the evolution equations
we derived in the previous sections for the twist-3 correlation
functions. First of all, we perform an extensive numerical study of
the exact Eq.\ (\ref{Mellin-Z}). For simplicity we restrict
ourselves to the homogeneous case, i.e.\  we discard
the quark-mass operator, which is certainly a justified assumption
for the light $u$- and $d$-quark species. The solution we are
interested in is given in terms of the eigenvalues and
eigenfunctions of the anomalous-dimension matrix
$_{\scriptscriptstyle ZZ}\gamma_n^l$ of the local operators
${\cal Z}_n^l$ calculated in Appendix C. The eigenvalue problem we
have attacked has no analytical solution; however, the
diagonalization can be done numerically for moderately
large orbital momentum $n$, e.g.\  $n\leq 100$, which is quite
sufficient for practical purposes. Second, we provide the analytical
solution of the generalized DGLAP equation in the multicolour
limit of QCD. In this case it reduces to the familiar ladder-type
equation that holds for the twist-2 operators. It will be shown
that such a reduction occurs in the limit $x \to 1$ too.
Assembling the two results allows us to construct the improved
DGLAP-type equations for two-quark twist-3 distributions. Exploiting
some examples of the evolution for the moments, we confront these two
approaches. In particular, we study the accuracy of the improved DGLAP
equation with respect to the exact evolution for different
models of gluon light-cone position distribution for the
three-particle correlation function at low momentum scale.

\subsection{Evolution of the moments}

To obtain the solution of the evolution equation (\ref{Mellin-Z}) we
choose $n$ as a positive integer. In this case, as follows from the
definition (\ref{def3part}) of ${\scriptstyle\cal Z}^n(u)$ the
$n$-th moment is actually given by the following linear combination
of local operators ${\cal Z}_n^l$ (see Eq. (\ref{defmom})):
\begin{eqnarray}
\label{Z(u)bylocalO}
{\scriptstyle\cal Z}^n(u)
=\sum_{l=1}^{n} C_{n-1}^{l-1}\,
u^{n-l} {\bar u}^{l-1} {\cal Z}_n^l,
\end{eqnarray}
so that ${\scriptstyle\cal Z}^n(u)$ is a polynomial of degree $n-1$
in $u$. Thus the kernel $P^n_{{\scriptscriptstyle\cal Z Z}}(u,v)$
possesses $n$ polynomial eigenfunctions $e^n_{l}(v)$:
\begin{eqnarray}
\int dv\,P^n_{{\scriptscriptstyle\cal Z Z}}(u,v) e^n_{l}(v)
= -\lambda^n_{l} e^n_{l}(u),
\quad l=1,\dots,n,
\end{eqnarray}
where $-\lambda^n_{l} $ denotes the eigenvalues. These eigenfunctions
can be constructed by diagonalization
\begin{eqnarray}
\label{reltolocanomdim}
C_{n-1}^{k-1} \int dv\,P^n_{{\scriptscriptstyle\cal Z Z}}(u,v)
v^{n-k} \bar{v}^{k-1} &=&
\sum_{l=1}^{n} C_{n-1}^{l-1}
{_{\scriptscriptstyle ZZ}\gamma}^n_{lk} \, u^{n-l} \bar{u}^{l-1},
\end{eqnarray}
where the anomalous-dimension matrix
${_{\scriptscriptstyle ZZ}\gamma}^n_{lk}$ of the local operators
is given by Eq. (\ref{anomdimZ}). Actually, this
is a purely algebraic task and we find
\begin{eqnarray}
\label{eigenfunc}
e^n_{k}(u)
= \sum_{l=1}^{n} C_{n-1}^{l-1}\,u^{n-l} \bar{u}^{l-1} E^n_{lk},
\mbox{\ with\ }
\left\{ (E^n)^{-1} {_{\scriptscriptstyle ZZ}\gamma}^n E^{n}
\right\}_{kl} = -\lambda^n_{k}\delta (k - l),
\end{eqnarray}
where $\delta (k - l)$ is a Kronecker symbol defined by Eq.\ 
(\ref{Kron}). The spectrum of the eigenvalues $\lambda^n_l$ up to
$n=50$ is shown in Fig.~\ref{spectrum}a, together with some
examples of the eigenfunctions of the kernel
$P^n_{\scriptscriptstyle\cal ZZ}$ for
particular orbital momenta. The solution for the moments
${\scriptstyle\cal Z} ^n(u)$ (in the massless case) is then
expressed in terms of the eigenfunctions and eigenvalues we have
found:
\begin{eqnarray}
\label{solution}
{\scriptstyle\cal Z}^n\left(u,Q^2\right) =
\sum_{l=1}^{n} c^n_l\left(Q_0^2\right) e^n_{l}(u)
\exp{ \left\{
-\int_{Q_0^2}^{Q^2} \frac{dt}{t} \frac{\alpha_s(t)}{2\pi}
\lambda^n_{l}
\right\}}.
\end{eqnarray}
The coefficients $c^n_l\left(Q_0^2\right)$ at the
reference momentum squared $Q^2_0$ have to be determined from the
non-perturbative input ${\scriptstyle\cal Z}^n\left(u,Q_0^2\right)$:
\begin{eqnarray}
\label{initialcondition}
c^n_l\left(Q_0^2\right) =
\sum_{k=1}^n (E^n_{lk})^{-1} \frac{(n-k)!}{(n-1)!}
\frac{d^{k-1}}{d w^{k-1}} (1+w)^{n-1}
{\scriptstyle\cal Z}^n\left({1\over 1+w},Q_0^2\right)_{|w=0}.
\end{eqnarray}

\subsection{Reduction of the evolution equations}

\begin{figure}[htb]
\hspace{-1cm}
\mbox{
\begin{picture}(170,110)

\put(0,0)	{
\epsffile{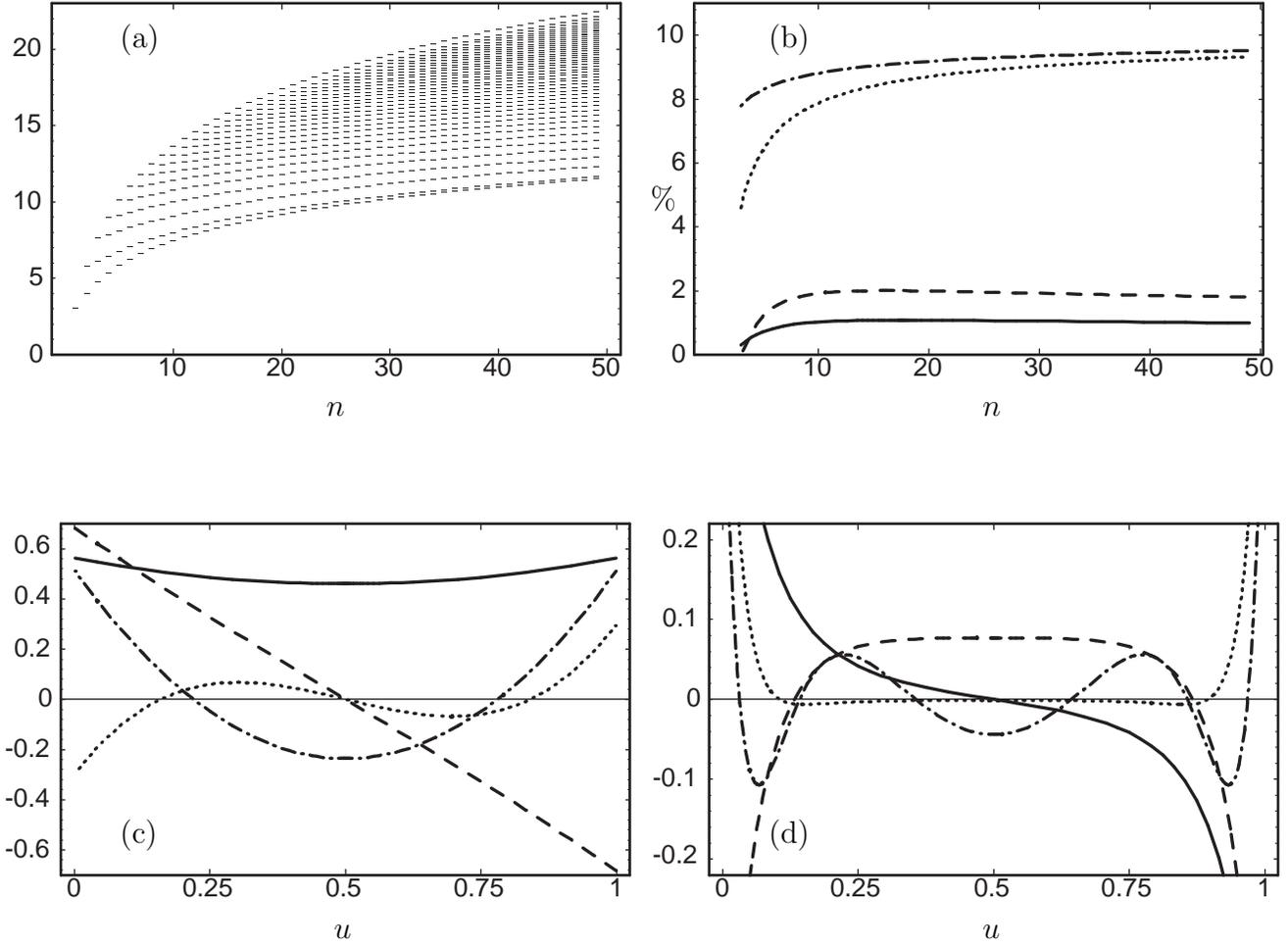}
		}
\put(23,122){(a)}
\put(51,72){$n$}
\put(111,122){(b)}
\put(140,72){$n$}
\put(23,14){(c)}
\put(95,100){\%}
\put(52,1){$u$}
\put(111,14){(d)}
\put(140,1){$u$}
\end{picture}
}
\vspace{2cm}
{\caption{\label{spectrum}
The spectrum of the eigenvalues $\lambda^n_l$ for the
evolution kernel $P^n_{\scriptscriptstyle\cal ZZ}$ defined in
(\ref{DefPZZ}) is shown in (a). In (b) the relative deviation
$1-\lambda^n_l/(-\gamma^n_l)$ (in \%) for the lowest two
eigenvalues of the spectrum, {\it i.e.} $l=1,2$, is plotted:
the solid (dashed) line $\gamma_n^1$ ($\gamma_n^2$) corresponds
to the $n$-th moments of the improved kernel $P_{ee}$
($P_{\bar h \bar h}$) defined by Eq. (\ref{imprKernel}); the
dash-dotted (dotted) line is the multicolour approximation for
$P_{ee}$ ($P_{\bar h \bar h}$). In the improved approximations,
subleading terms were taken into account to reproduce the first
two eigenvalues exactly. Eigenfunctions of the kernel
$P^n_{\cal ZZ}$ are shown for $n=4$ in (c) and for $n=30$ with
$l=2,3,6,29$ in (d).
}}
\end{figure}

In the large-$N_c$ limit only the planar diagrams (Fig. 1 a, d)
survive and the kernel $P^n_{\scriptscriptstyle\cal Z Z}(u,v)$ has
two known dual eigenfunctions: $1$ and $1 - 2u$, so that
$\int_0^1 du\, e^n_{l}(u) = \delta_{l1} + {\cal O}(1/N_c)$ and
$\int_0^1 du\, (1-2u) e^n_{l}(u) = \delta_{l2} + {\cal O}(1/N_c)$,
where $l=1,2$ correspond to the lowest two eigenvalues of the
spectrum shown in Fig.~\ref{spectrum}a. A straightforward
calculation gives the following DGLAP evolution kernels:
\begin{equation}
\label{largeNc-ZZ}
\int_0^1 du\,
\!\left\{ \!
\begin{array}{c}
1 \\
\frac{1-2u}{1-2v}
\end{array}
\!\right\}
P_{\scriptscriptstyle\cal Z Z}(x,u,v)
= N_c\,\theta (\bar x)\theta (x)
\!\left\{ \!
\begin{array}{c}
\left[ \frac{x^2}{\bar x} \right]_+ + \frac{1}{2}x^2
- \frac{5}{4} \delta(\bar x) \\
\left[ \frac{x^2}{\bar x} \right]_+ - \frac{3}{2} x^2
- \frac{5}{4} \delta(\bar x)
\end{array}
\!\right\} + {\cal O}\left(\frac{1}{N_c}\right),
\end{equation}
and for the mass-mixing kernels the exact results read
\begin{eqnarray}
&&\int_0^1 du\,
\!\left\{ \!
\begin{array}{c}
1 \\
\frac{1-2u}{1-2v}
\end{array}
\!\right\}
P_{{\scriptscriptstyle\cal Z} m}(x,u,v)
=
C_F\, \theta (\bar x)\theta (x)
\!\left\{ \!
\begin{array}{c}
x (2 - x)\\
0
\end{array}
\!\right\}
\quad\mbox{for e},\nonumber\\
&&\int_0^1 du
\!\left\{ \!
\begin{array}{c}
1 \\
\frac{1-2u}{1-2v}
\end{array}
\!\right\}
P_{{\scriptscriptstyle\cal Z} m}(x,u,v)
=
C_F\, \theta(\bar x) \theta (x)
\!\left\{ \!
\begin{array}{c}
0 \\
x (2 - 3 x)
\end{array}
\!\right\}\quad\mbox{for $\widetilde{h}_L$}.
\end{eqnarray}
As was first observed in Ref. \cite{ali91} in the context
of the chiral-even distribution $g_2(x)$, similar equations hold
true also for the $\frac{1}{N_c}$-suppressed terms in the $x\to 1$
limit for flavour non-singlet twist-3 evolution kernels. In the
present chiral-odd case, we find
\begin{eqnarray}
\label{largeNc-ZZ-impr}
\int_0^1 du\,
\!\left\{ \!
\begin{array}{c}
1 \\
\frac{1-2u}{1-2v}
\end{array}
\!\right\}
P_{\scriptscriptstyle\cal Z Z}(x,u,v)
=
-\frac{1}{N_c} \theta(\bar x) \theta (x) \!\left\{ \!
\begin{array}{c}
\left[\frac{1}{\bar{x}}\right]_+ + \frac{5}{4}\delta(\bar{x}) +
{\cal O}(\bar{x}^0)
\\
\left[\frac{1}{\bar{x}}\right]_+ + \frac{19}{12}\delta(\bar{x})+
{\cal O}(\bar{x}^0)
\end{array}
\!\right\} + N_c \cdots,
\end{eqnarray}
where the $ N_c \cdots $ symbolize the $x \to 1$ limit of
Eq. (\ref{largeNc-ZZ}).

The eigenfunctions we have obtained coincide precisely with the
coefficients that appear in the decomposition of $e(x,Q^2)$ and
$\widetilde{h}_L(x,Q^2)$ in terms of three-particle correlation
functions. To observe this explicitly we need relations similar
to the ones given by Eqs. (\ref{EOM-e}) and (\ref{sumrule}) transformed
to the mixed representation (\ref{def3part}). Namely, we have
\begin{eqnarray}
\label{rel-to-e}
e(x) &=&
\frac{1}{x} m(x) -
\frac{1}{2}\frac{d}{dx} \int_0^1du\, {\scriptstyle\cal Z}(x,u), \\
\label{rel-to-h}
\bar{h}_L(x)&=&
\frac{1}{x} \widetilde m(x) -
\frac{1}{2}\frac{d}{dx} \int_0^1du\,
(1-2u)\widetilde {\scriptstyle\cal Z}(x,u),
\end{eqnarray}
where we introduce for convenience a new function $\bar{h}_L(x)$,
so that ${h}_L(x)$ reads:
\begin{eqnarray}
\label{sumrule2}
{h}_L(x)=
2 \int_x^1dy\, \frac{x^2}{y^2} h_1(y) -
x\frac{d}{dx} \int_x^1 \frac{dy}{y} \frac{x^2}{y^2} \bar{h}_L(y).
\end{eqnarray}
For the massless case the last term on the RHS coincides with
the twist-3 part $\widetilde{h}_L$.

From the observations we have made above, it follows that
in the large-$N_c$ as well as in the large-$x$ limit the twist-3
distributions satisfy the DGLAP evolution equations. By combining
the large-$N_c$ evolution with the large-$x$ result for
the $\frac{1}{N_c}$-suppressed terms, we can improve the accuracy of
such an approximation within a factor 5--10, to be compared with the
multicolour limit taken alone (see Fig.~\ref{spectrum}b).
Thus, the functions $e(x,Q^2)$ and ${\bar h}_L(x,Q^2)$ obey the
following improved evolution equations:
\begin{eqnarray}
\label{DGLAPforchiodd}
\left\{\begin{array}{c}
\dot e(x) \\
{\dot {\bar h}}_L(x)
\end{array}\right\}
= {\alpha_s \over 2\pi} \int_x^1 \frac{dy}{y}
\left\{\begin{array}{c}
P_{ee}(y)\, e\left( \frac{x}{y} \right) +
\left[ 	\frac{1}{x}\left\{P_{mm}(y)-P_{ee}(y)\right\}
-\frac{1}{2}P_{em}(y)\frac{d}{dx} \right] m\left( \frac{x}{y}
\right) \\
P_{\bar h\bar h}(y)\,{\bar h}_L \left( \frac{x}{y} \right) +
\left[ \frac{1}{x}\left\{P_{mm}(y)-P_{\bar h\bar h}(y)\right\}-
\frac{1}{2}P_{\bar h m}(y)\frac{d}{dx}
\right] m\left( \frac{x}{y} \right)
\end{array}\right\} ,
\end{eqnarray}
with
\begin{eqnarray}
\label{imprKernel}
&&P_{ee}(y) =
2C_F \left[\frac{y}{\bar y}\right]_+ + \frac{C_A}{2}y +
\left(\frac{C_F}{2}-C_A\right)  \delta(\bar y) +
{\cal O}\left({\bar y}^0/N_c\right),\nonumber\\
&&P_{\bar h \bar h}(y)
= 2C_F \left[\frac{y}{\bar y}\right]_+
- \frac{3C_A}{2}y +\left(\frac{7C_F}{6}
-\frac{4C_A}{3}\right)\delta(\bar y)
+{\cal O}\left({\bar y}^0/N_c\right),\\
&&P_{em}(y)
= C_F
x (2 - x),\nonumber\\
&&P_{\bar h m}(y)
= C_F x (2 - 3 x).
\end{eqnarray}
The evolution kernel for the mass-dependent correlator was already
given by Eq. (\ref{Pmm}). In Eq. (\ref{imprKernel}) we have added
subleading terms  ${\cal O}\left({\bar y}^0/N_c\right)$
(which we do not specify here) so that the first moment of each
kernel coincides with the corresponding eigenvalue of the kernel
$P^n_{{\scriptscriptstyle\cal Z Z}}(u,v)$. This guarantees that
the solution for the lowest moments given below, in Eq.\ 
(\ref{evol-low-moments}), will be reproduced exactly. Note that
in the massless case $\widetilde{{h}}_L(x)$ fulfills the same
evolution equation as ${\bar h}_L(x)$. The simplest way to
verify this is to make the Mellin transform of the corresponding
evolution equations.

Let us add a few remarks on the momentum space formulation. As we
have seen above the solution of the evolution equations in the
asymptotic regimes is the most straightforward in the light-cone
position representation. It is by no means trivial to observe
the appearance of the DGLAP equations in momentum fraction
representation. However, we know that the asymptotic solution, in
coordinate space, is given by
the convolution of the three-particle correlation function with
the same weight function that enters in the decomposition of the
two-parton correlators at tree level.
With this in mind, we are able to check that the integrals
\begin{eqnarray}
e(x) &=& \int d\beta' D(x, \beta'),\\
\widetilde h_L (x) &=& x^2 \int_{x}^{1} \frac{d\beta}{\beta^2}
\int \frac{d\beta'}{\beta' - \beta}
\left\{ 2+
(\beta - \beta')
\left[ \frac{\partial}{\partial \beta}
- \frac{\partial}{\partial \beta'} \right]
\right\}
\widetilde D (\beta, \beta'),
\end{eqnarray}
taken from Eqs. (\ref{EOM-e}) and (\ref{sumrule}) neglecting quark-mass
as well as twist-2 effects, satisfy the DGLAP equations,
namely
\begin{eqnarray}
&&\hspace{-1.3cm}\dot e(x) \nonumber\\
&&\hspace{-1.3cm}= - \frac{\alpha}{4\pi} N_c \int d \beta e(\beta)
\left\{
2
\left[
\frac{ \beta }{( x - \beta )}\Theta_{11}^0 (x, x - \beta)
\right]_+
+
\left( 2 - \frac{x}{\beta} \right) \Theta_{11}^0 (x, x-\beta)
- \frac{1}{2} \delta (\beta - x)
\right\}, \\
&&\hspace{-1.3cm}\dot{\widetilde h}_L(x)\nonumber\\
&&\hspace{-1.3cm}= - \frac{\alpha}{4\pi} N_c \int d \beta
\widetilde h_L(\beta)
\left\{
2
\left[
\frac{ \beta }{( x - \beta )}\Theta_{11}^0 (x, x - \beta)
\right]_+
+
\left( 2 + 3 \frac{x}{\beta} \right) \Theta_{11}^0 (x, x-\beta)
- \frac{1}{2} \delta (\beta - x)
\right\}.
\end{eqnarray}
The corresponding anomalous dimensions are
\begin{eqnarray}
\dot{[e]}_n
&=& \frac{\alpha}{4\pi} N_c
\left\{
-2\psi (n+2) - 2\gamma_E + \frac{1}{2} + \frac{1}{n+2}
\right\}
[e]_n , \\
\dot{[\widetilde h_L]}_n
&=& \frac{\alpha}{4\pi} N_c
\left\{
-2\psi (n+2) - 2\gamma_E + \frac{1}{2} - \frac{3}{n+2}
\right\}
[\widetilde h_L]_n.
\end{eqnarray}
Which are exactly the anomalous dimensions $\gamma^\pm_n$ found in
Ref. \cite{bbkt96} for $e$ and $h_L$, respectively, with the replacement
$n \to j-1$ \footnote{The difference in the anomalous dimensions is
due to an extra power of the momentum fraction $x$ included in the
definition of the twist-3 correlation functions.}.

\subsection{Examples of the evolution}
\label{numstudies}

For the lowest few moments the evolution equation (\ref{Mellin-Z})
can be solved exactly. Taking into account the symmetry properties of
the quark--gluon correlation function, we obtain the following results
for the (non-vanishing) first two moments:
\begin{eqnarray}
\label{evol-low-moments}
{\scriptstyle\cal Z}^1\left(u,Q^2\right) &=&
{\scriptstyle\cal Z}^1\left(Q_0^2\right)
\exp\left\{ -\frac{55}{18}
\int_{Q_0^2}^{Q^2} \frac{dt}{t} \frac{\alpha_s(t)}{2\pi}\right\},
\nonumber\\
{\scriptstyle\cal Z}^2\left(u,Q^2\right) &=&
{\scriptstyle\cal Z}^2\left(Q_0^2\right)
\exp\left\{ -\frac{73}{18}
\int_{Q_0^2}^{Q^2} \frac{dt}{t} \frac{\alpha_s(t)}{2\pi}\right\}
\nonumber\\
\widetilde{\scriptstyle\cal Z}^2\left(u,Q^2\right) &=&
\widetilde{\scriptstyle\cal Z}^2\left(Q_0^2\right)
(1-2u) \exp\left\{ -\frac{52}{9}
\int_{Q_0^2}^{Q^2} \frac{dt}{t} \frac{\alpha_s(t)}{2\pi}\right\},
\nonumber\\
\widetilde{\scriptstyle\cal Z}^3\left(u,Q^2\right) &=&
\widetilde{\scriptstyle\cal Z}^3\left(Q_0^2\right)
(1-2u) \exp\left\{ -\frac{1099}{180}
\int_{Q_0^2}^{Q^2} \frac{dt}{t} \frac{\alpha_s(t)}{2\pi}\right\},
\end{eqnarray}
where ${\scriptstyle\cal Z}^n\left(Q_0^2\right)$ are related to the
following matrix elements of the local operators (up to
normalization)
\begin{eqnarray}
\label{initcond}
{\scriptstyle\cal Z}^1\left(Q_0^2\right)
&=& Z^1_1\left(Q_0^2\right)
= \langle h |{\cal  Z}^1_1 | h \rangle_{|\mu^2=Q_0^2},
\quad
{\scriptstyle\cal Z}^2\left(Q_0^2\right)
=  Z^1_2\left(Q_0^2\right)
= \langle h |{\cal  Z}^1_2 | h \rangle_{|\mu^2=Q_0^2},
\nonumber\\
\widetilde {\scriptstyle\cal Z}^2\left(Q_0^2\right)
&=& \widetilde Z^1_2\left(Q_0^2\right)
= \langle h |\widetilde {\cal  Z}^1_2 | h \rangle_{|\mu^2=Q_0^2},
\quad
\widetilde {\scriptstyle\cal Z}^3\left(Q_0^2\right)
= \widetilde Z^1_3\left(Q_0^2\right)
= \langle h |\widetilde{\cal  Z}^1_3 | h \rangle_{|\mu^2=Q_0^2}.
\end{eqnarray}

\begin{figure}[htb]
\hspace{-1cm}
\mbox{
\begin{picture}(170,110)

\put(0,0)	{
\epsffile{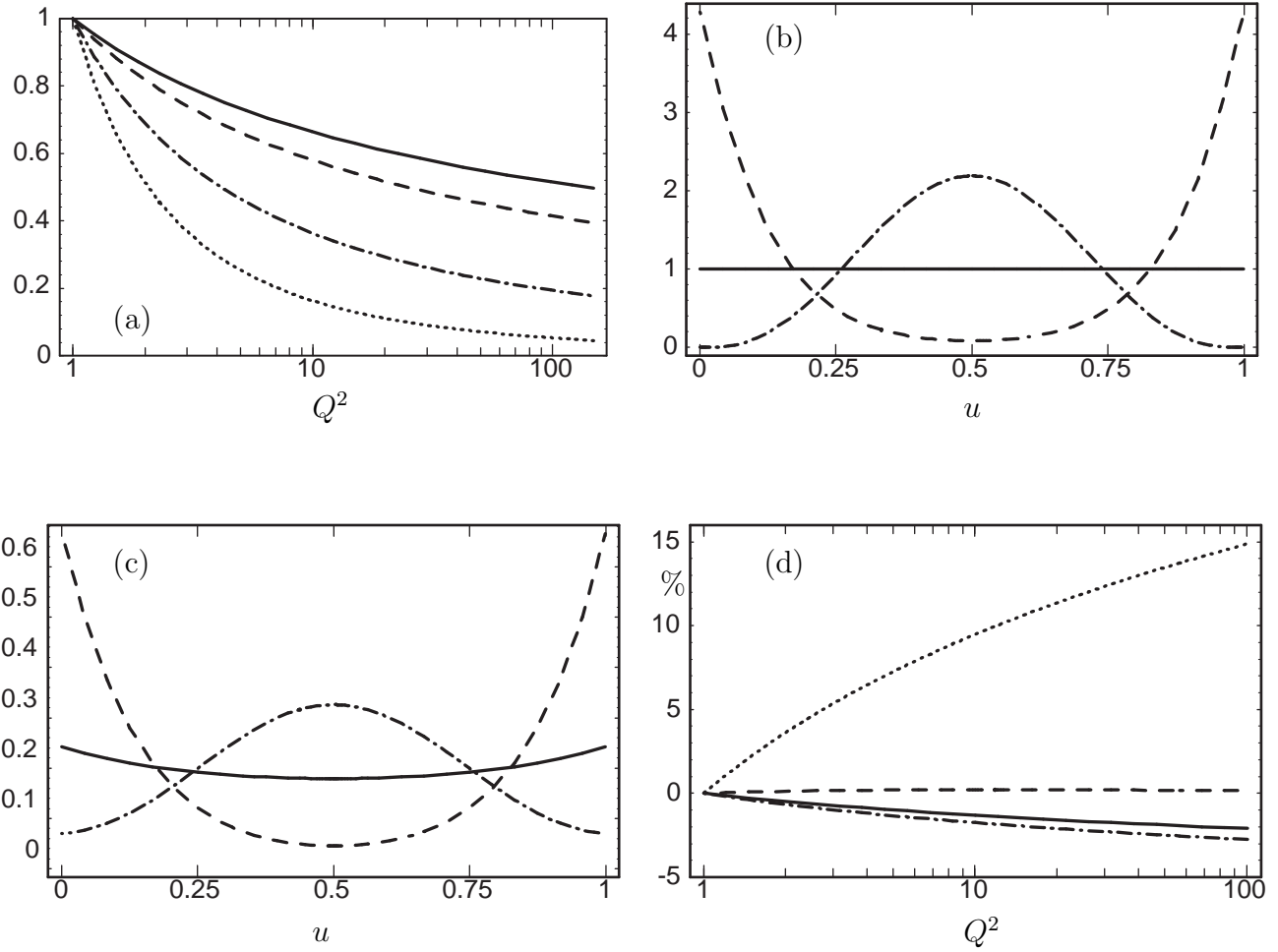}
		}
\put(23,84){(a)}
\put(50,72){$Q^2$}
\put(111,122){(b)}
\put(138,72){$u$}
\put(23,51){(c)}
\put(50,1){$u$}
\put(111,51){(d)}
\put(138,1){$Q^2$}
\put(97,48){\%}
\end{picture}
}
\vspace{1.5cm}
\caption{\label{evol-e}
The predictions of the improved evolution equation for the moments
$[e]_n$, normalized to 1 at $Q_0^2=1\mbox{\ } {\rm GeV}^2$, are shown
in (a) for $n=1$ (solid line), $n=2$ (dashed line), $n=10$ (dash-dotted
line) and $n=100$ (dotted line). In (b) three different models for
the gluon light-cone position distributions at $Q_0^2=1\mbox{\ }
\rm{GeV}^2$ for $n=10$ are presented; they are evolved up to the scale
$Q^2=100\mbox{\ }{\rm GeV}^2$ in (c). The relative deviation
$1-[e(Q^2)]_n^{\rm ex}/[e(Q^2)]_n^{\rm im}$ (in \%) from the improved
DGLAP equation of the exact evolution for the assumed gluon distributions
is shown in (d). The dotted line corresponds to the
relative deviation $1-[e(Q^2)]_n^{N_c}/[e(Q^2)]_n^{\rm coef}$ with
respect to the multicolour approximation, where ``coef" refers to the
gluon distribution that is equivalent to the corresponding
coefficient function.
}
\end{figure}

\begin{figure}[htb]
\hspace{-1cm}
\mbox{
\begin{picture}(170,110)

\put(0,0)	{
\epsffile{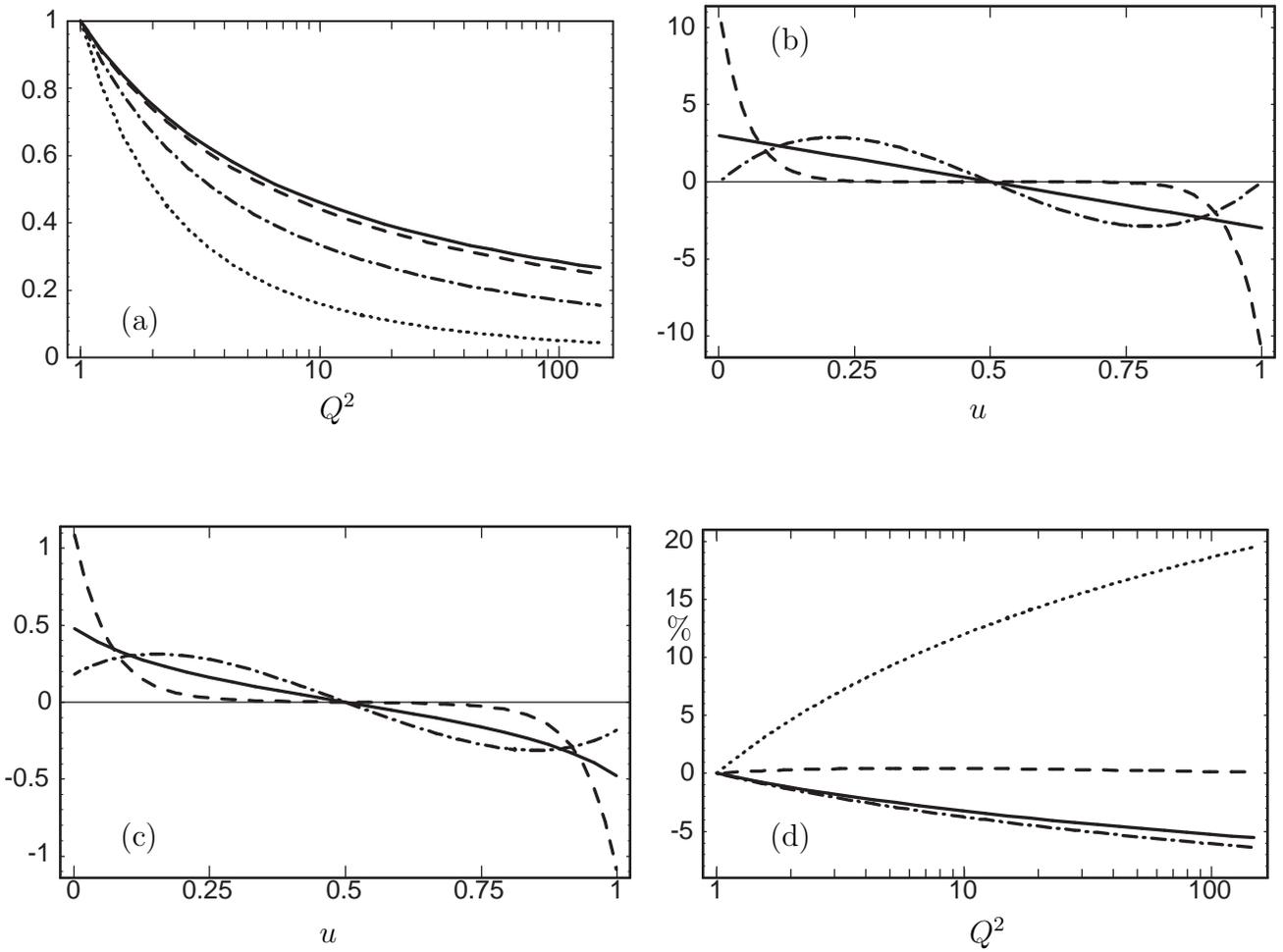}
		}
\put(23,84){(a)}
\put(50,72){$Q^2$}
\put(111,122){(b)}
\put(138,72){$u$}
\put(23,14){(c)}
\put(50,1){$u$}
\put(111,14){(d)}
\put(138,1){$Q^2$}
\put(97,42){\%}
\end{picture}
}
\vspace{1.5cm}
\caption{\label{evol-h}
The predictions of the improved evolution equation for the moments
$[\widetilde{h}_L]_n$, normalized to 1 at $Q_0^2=1\mbox{}\rm{GeV}^2$,
are shown in (a). Here the solid line and dashed line represent $n=2$
and $n=3$, respectively. The further description is the same as in
Fig.~\ref{evol-e}, except that $n=20$ in (b)--(d).
}
\end{figure}

Thus the $Q^2$-dependence for the first moments of $e(x)$ and
$\widetilde h_L(x)$ can be predicted uniquely since the initial
values at the low-momentum scale are given by these moments
themselves. For larger $n$, the evolution is sensitive to the shape
of the gluon distribution between the quark fields. However, as can
be traced from Figs.~\ref{evol-e} and \ref{evol-h}, the dependence
on the assumed different toy models of the light-cone position
distribution at $Q_0^2=1\mbox{\ }{\rm GeV}^2$ is quite small: the
typical relative deviations for the moments of the twist-3 unpolarized
and polarized structure functions are $2\%$ and $5\%$, respectively,
at $Q^2=100\mbox{\ }{\rm GeV}^2$. In the calculations we have set the
number of flavours $N_f=3$ and $\Lambda_{QCD}=0.25\mbox{\ }{\rm GeV}$.
For the ``gap"-type (end-point-concentrated) distributions
(see Figs.~\ref{evol-e}, \ref{evol-h} dashed line) the deviation is
small with respect to the ``coefficient function"-type model
predictions ($1$ and $1-2u$, solid lines), while for the ``hump"
(end-point-suppressed) distributions (dash-dotted line) it is
a little bit larger. The accuracy of the multicolour approximation
is about 15--20\% at a scale $Q^2=100\mbox{\ }{\rm GeV}^2$. These numbers
are natural and can be expected from the discrepancy between the
DGLAP anomalous dimensions and the exact lowest two eigenvalues of
the spectrum (see Fig.~\ref{spectrum}b). Nevertheless, there is
one very important exception from the naive expectation. If the
initial gluon distribution is strongly suppressed in the end-point
region, for instance as $u^{[(n-1)/2]}$ for $u \to 0$, then the
large-$N_c$ approximation breaks down for large $n$ (polarized
moments are more sensitive than unpolarized ones). In this case
the evolution is not smooth. The shape of this function will
be turned immediately into the end-point-concentrated one. So, we
can get rid of this ``hump"-type model for a momentum transfer
$Q \stackrel{\textstyle >}{\sim} 1$ GeV and argue that such a
distribution could not occur in the non-perturbative domain either.
It is the most likely to assume that the momentum fraction-function
$Z(x,x')$ is rather smooth. From the equation
\begin{eqnarray}
\label{relatMoments}
{\scriptstyle\cal Z}^n(u)
= \int dx\int dx' (x u + x' \bar{u})^{n-1}\, Z(x,x')
\end{eqnarray}
it then follows that ${\scriptstyle\cal Z}^n(u)$ cannot be strongly
suppressed in the end-point region. For instance, if $Z(x,x')$ is
positive-definite and concentrated in the region $0\le x,x'$, then
${\scriptstyle\cal Z}^n(u)$ cannot vanish at $u=0,1$ unless it
is identically zero.

\section{Discussion and conclusion}

In the present paper we have investigated the $Q^2$-dependence of
the chiral-odd distributions of the nucleon $e(x)$ and $h_L(x)$.
Using the constraint equalities coming from the equation of motion
and Lorentz invariance, which provide  certain sum rules for
the structure function, the problem is reduced to a study of the
renormalization of the multiparton correlators in lowest
order of the perturbation theory. As a result we construct
an exact (taking account of $1/N^2_c$ effects) one-loop evolution
in the light-cone fraction as well as in the light-cone position
representations. For these purposes, we have used two techniques,
which employ the light-like gauge for the gluon field. Accepting
different prescriptions on the spurious pole in the gluon propagator,
we were able to verify that they do lead to the same results. From
the calculational point of view the momentum space technique is much
easier to treat. However, the coordinate space makes the involved
symmetries apparent and, as a by-product, diagonalization of
evolution kernels is easy to handle. We establish the bridge between
different formulations of the QCD evolution. It is straightforward to
obtain the evolution kernels in the light-cone fraction representation,
starting from the coordinate space and vice versa using the Fourier
transform. Using this transformation it is straightforward to
obtain the coordinate space two-particle evolution kernels for
Faddeev-type equations with pair-wise particle interaction, which
govern the scale dependence for higher twist (more than 3)
quasi-partonic correlators \cite{lip85} and to study the
diagonalization problem, which is more easier to do in the light-cone
position representation. We hope to return to this question in the
future.

In the multicolour limit as well as for $x \to 1$ we obtain the
ladder-type evolution equations for the twist-3 part of the
distribution functions. Joining these asymptotics together we
construct an improved DGLAP equation, which generally has very good
accuracy, at the level of few per cent. We argue that the observed
discrepancy between these reduced equations and the exact evolution
could occur only for unphysical initial conditions of the
latter. To clarify the situation completely, it would be helpful to
have the low-energy model predictions for the distribution of the
gluon field in the quark--gluon correlator.

\bigskip
{\bf Acknowledgements.} We would like to thank V.M. Braun for
discussions at an early stage of the work. The authors are grateful
to the CERN Theory Division for its hospitality during their visit,
where this work was started. A.B. was supported by the Russian
Foundation for Fundamental Research, grant N 96-02-17631. D.M. was
financially supported by the Deutsche Forschungsgemeinschaft (DFG).

\appendix

\setcounter{section}{0}
\setcounter{equation}{0}
\renewcommand{\theequation}{\Alph{section}.\arabic{equation}}

\section{Definition and some properties of $\Theta$-functions}

The $\Theta$-functions entering the evolution equations in
the momentum fraction representation are given by the formula
\begin{equation}
\Theta^{m}_{i_1 i_2 ... i_n}
(x_1,x_2,...,x_n)=\int_{-\infty}^{\infty}\frac{d\alpha}{2\pi i}
\alpha^m \prod_{k=1}^{n}\left(\alpha x_k -1 +i0 \right)^{-i_k}.
\end{equation}
For our practical purposes it is enough to have an explicit form
of the functions
\begin{eqnarray}
&&\Theta^0_1 (x) = 0,\\
&&\Theta^0_2 (x) = \delta (x),\\
&&\Theta^0_{11} (x_1,x_2)
=\frac{\theta(x_1)\theta(-x_2)
-\theta(x_2)\theta(-x_1)}{x_1-x_2},
\end{eqnarray}
since the other are expressed in their terms by the relations
\begin{eqnarray}
&&\Theta^0_{21} (x_1,x_2)
=\frac{x_2}{x_1-x_2}\Theta^0_{11} (x_1,x_2),\\
&&\Theta^1_{21} (x_1,x_2)
=\frac{1}{x_1-x_2}\Theta^0_{11} (x_1,x_2)
- \frac{1}{x_1-x_2}\Theta^0_2 (x_1),\\
&&\Theta^0_{22} (x_1,x_2)
= -\frac{2x_1x_2}{(x_1-x_2)^2} \Theta^0_{11} (x_1,x_2),\\
&&\Theta^0_{111} (x_1,x_2,x_3)
=\frac{x_2}{x_1-x_2}\Theta^0_{11} (x_2,x_3)
-\frac{x_1}{x_1-x_2}\Theta^0_{11} (x_1,x_3),\\
&&\Theta^1_{111} (x_1,x_2,x_3)
=\frac{1}{x_1-x_2}\Theta^0_{11} (x_2,x_3)
-\frac{1}{x_1-x_2}\Theta^0_{11} (x_1,x_3).
\end{eqnarray}
In the main text we have used the relations
\begin{equation}
\int d\beta \beta^n
\Theta^0_{11} (\beta, \beta - x')
\Theta^0_{11} (x, x - \beta)
= x'^n \Xi_n (x, x - x', x)
=\frac{1}{n}\left[
x'^n - x^n \right]
\Theta^0_{11} (x, x-x').
\end{equation}
\begin{equation}
{\rm PV}\int d\beta \frac{x}{(x - \beta)}
\left[
\Theta^0_{11} ( \beta , \beta - x)
+\Theta^0_{11}(x, x - \beta)
\right]=0.
\end{equation}

\setcounter{equation}{0}

\section{Evolution equations in Abelian gauge theory}

In this appendix we present a pedagogical illustration of the
renormalization group mixing problem for the redundant basis of
correlation functions defined by Eqs.
(\ref{DefUnpoCF-e})--(\ref{DefUnpoCF-D2}) and Eqs.
(\ref{DefPoCF-h_1})--(\ref{DefPoCF-D2})
for the unpolarized and polarized cases, respectively,
in the framework of the Abelian gauge theory. Its aim is to
show the self-consistency of the whole approach we have used
as the equations derived below satisfy the constraint equalities
given by Eqs. (\ref{EOM-e}), (\ref{EOM-h_L}), (\ref{tw-2-3}), which
are further employed to reduce the overcomplete set of correlators
to the independent basis of functions.

\begin{figure}[htb]
\mbox{
\hspace{3cm}
\epsffile{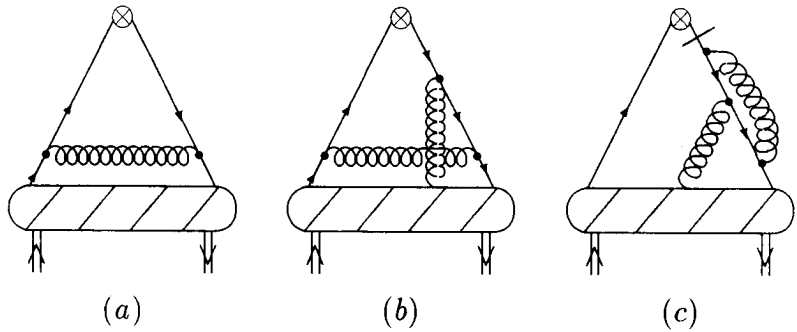}}
\vspace{0.1cm}
{\caption{\label{two-part}
One-loop radiative corrections for two-particle
correlators in the Abelian gauge theory. The fermion propagator
crossed with a bar on diagram (c) shows the contraction of the
corresponding line into the point.
}}
\end{figure}

In the calculations of the corresponding evolution kernels, we
follow the methods developed in Ref. \cite{lip84} (see Ref.
\cite{bel96} for a more recent discussion of the RG equations
for the time-like twist-3 cut vertices). The
one-loop Feynman diagrams giving rise to the transition
amplitudes of two-particle correlation functions into the two-
and three-parton ones are shown in Fig.~\ref{two-part}a, b.
The last figure (c) on this picture is specific of the vertices
having non-quasi-partonic form \cite{lip85}, that is for $e(x)$
and $h_L(x)$; it displays the addendum due to the contact term
that results from the cancellation of the propagator adjacent to
the quark--gluon and bare vertices. As an output the vertex acquires
the three-particle piece. The radiative correction to the
three-parton correlators are presented in Fig.~\ref{three-pa}
a, b, c.

A straightforward calculation yields the evolution equations
for the spin-independent case in the form
\begin{eqnarray}
\dot M(x)
&=& - \frac{\alpha}{2\pi}\int d \beta M(\beta)
\left\{
2
\left[
\frac{ \beta }{( x - \beta )}\Theta_{11}^0 (x, x - \beta)
\right]_+
+
\frac{\beta + x}{\beta} \Theta_{11}^0 (x, x-\beta)
\right\},\\
&&\nonumber\\
\dot e(x)
&=&\frac{\alpha}{2\pi}\int d\beta
\Biggl(
e (\beta )
\left\{
\frac{x}{\beta}\Theta_{11}^0 (x, x - \beta)
+\frac{1}{2} \delta (\beta - x)
\right\} \nonumber\\
&-&
M (\beta)
\left\{
2
\left[
\frac{ \beta }{( x - \beta )}
\Theta_{11}^0 (x, x - \beta)\right]_+
+
x \Theta_{21}^1 (x, x - \beta)
+
2 \Theta_{11}^0 (x, x - \beta)
\right\}
\nonumber\\
&-&\int d\beta' D(\beta , \beta')
\Biggl\{
2
\left[
\frac{ \beta }{(x - \beta)}
 \Theta_{11}^0 (x, x - \beta)
\right]_+
+ \frac{x}{x - \beta}
\Theta_{111}^0 (x , x - \beta, x - \beta + \beta') \nonumber\\
&+&\delta (\beta - x) \int d \beta '' \frac{\beta}{\beta''}
\Theta_{111}^0 (\beta'',\beta'' - \beta, \beta'' - \beta')
+
2 \Theta_{11}^0 (x, x - \beta)
\Biggr\}
\Biggr),\\
&&\nonumber\\
\dot D (x , x')&=&
-\frac{\alpha}{2\pi}
\Biggl\{
\left[
\frac{x'}{x}e(x) - M (x)
\right]
\Theta_{11}^0 (x', x' - x)
-
\left[
\frac{x}{x'}e(x') - M (x')
\right]
\Theta_{11}^0 (x, x - x')
\nonumber\\
&+&\int d\beta'
\Biggl(
D(x, \beta')\frac{(\beta' - x + x')}{(x - x')}
\Theta_{111}^0 (x', x' - x , x' - x + \beta')\nonumber\\
&+&\frac{x'}{x' - \beta'}
[D (x - x' + \beta', \beta') - D (x, x')]
\Theta_{11}^0 (x' , x' - \beta')
\Biggr) \nonumber\\
&+&\int d\beta
\Biggl(
D(\beta , x')\frac{(\beta - x' + x)}{(x' - x)}
\Theta_{111}^0 (x, x - x' , x - x' + \beta)\nonumber\\
&+&\frac{x}{x - \beta}
[D (\beta , x' - x + \beta) - D (x, x')]
\Theta_{11}^0 (x , x - \beta)
\Biggr) - \frac{3}{2}D(x,x')
\Biggr\},
\end{eqnarray}
where the dot denotes the derivative with respect to the UV cutoff
$\hspace{0.1cm}\dot{}=\Lambda^2{\partial}/{\partial\Lambda^2}$ and
the plus-prescription is defined by the equation
\begin{equation}
\label{plus}
\left[ \frac{\beta}{( x - \beta )}
\Theta_{11}^0 (x, x - \beta) \right]_+
=
\frac{ \beta }{( x - \beta )}
\Theta_{11}^0 (x, x - \beta)
-
\delta (\beta - x)
\int d \beta''
\frac{\beta}{ (\beta'' - \beta ) }
\Theta_{11}^0 ( \beta'', \beta'' - \beta).
\end{equation}

In the same way we can immediately obtain the set of evolution
equations for the spin-dependent distributions
(\ref{DefPoCF-h_1})--(\ref{DefPoCF-D2}):
\begin{eqnarray}
\dot{\widetilde M}(x)
&=& - \frac{\alpha}{2\pi}\int d \beta \widetilde M(\beta)
\left\{
2
\left[
\frac{ \beta }{( x - \beta )}\Theta_{11}^0 (x, x - \beta)
\right]_+
+
\frac{\beta + x}{\beta} \Theta_{11}^0 (x, x-\beta)
\right\},\\
&&\nonumber\\
\dot h_1(x)
&=& - \frac{\alpha}{2\pi}\int d \beta h_1(\beta)
\left\{
2
\left[
\frac{ \beta }{( x - \beta )}\Theta_{11}^0 (x, x - \beta)
\right]_+
+
2 \Theta_{11}^0 (x, x-\beta)
-\frac{3}{2} \delta (x - \beta)
\right\}\!,\nonumber\\
&&\\
\dot h_L(x)
&=&\frac{\alpha}{2\pi}\int d\beta
\Biggl(
h_L (\beta )
\left\{
\frac{x}{\beta}\Theta_{11}^0 (x, x - \beta)
+\frac{1}{2} \delta (\beta - x)
\right\} \nonumber\\
&-&
\widetilde M (\beta)
\left\{
2
\left[
\frac{ \beta }{( x - \beta )}
\Theta_{11}^0 (x, x - \beta)\right]_+
+
\left(
2 + \frac{x}{\beta}
\right)
\Theta_{11}^0 (x, x - \beta)
\right\}
\nonumber\\
&-&
K (\beta)
\left\{
2
\left[
\frac{ \beta }{( x - \beta )}
\Theta_{11}^0 (x, x - \beta)\right]_+
+
\left(
2 + 3\frac{x}{\beta}
\right)
\Theta_{11}^0 (x, x - \beta)
- \delta (x - \beta)
\right\}
\nonumber\\
&-&\int d\beta' \widetilde D(\beta , \beta')
\Biggl\{
2
\left[
\frac{ \beta }{(x - \beta)}
\Theta_{11}^0 (x, x - \beta)
\right]_+
+ \frac{x}{x - \beta}
\Theta_{111}^0 (x , x - \beta, x - \beta + \beta') \nonumber\\
&+&\delta (\beta - x) \int d \beta '' \frac{\beta}{\beta''}
\Theta_{111}^0 (\beta'',\beta'' - \beta, \beta'' - \beta')
+
2
\left(
1 + \frac{x}{\beta}
\right)
\Theta_{11}^0 (x, x - \beta)
\Biggr\}
\Biggr),\\
&&\nonumber\\
\dot K(x)
&=&\frac{\alpha}{2\pi}\int d\beta
\Biggl(
2 h_L (\beta )
\frac{x}{\beta}\Theta_{11}^0 (x, x - \beta)
- 2 \widetilde M (\beta)
\Theta_{11}^0 (x, x - \beta)
\\
&-&
K (\beta)
\left\{
2
\left[
\frac{ \beta }{( x - \beta )}
\Theta_{11}^0 (x, x - \beta)\right]_+
+
2 \left(
1 + 2 \frac{x}{\beta}
\right)
\Theta_{11}^0 (x, x - \beta)
-\frac{3}{2} \delta (x - \beta)
\right\}
\nonumber\\
&-&\int d\beta' \widetilde D(\beta , \beta')
\Biggl\{
2
\frac{(x - \beta + \beta' )}{(x - \beta)}
\Theta_{111}^0 (x, x - \beta, x - \beta + \beta' )
+ 2\frac{x}{ \beta}
\Theta_{11}^0 (x , x - \beta)
\Biggr\}
\Biggr),\nonumber\\
&&\nonumber\\
\dot{\widetilde D} (x , x')&=&
-\frac{\alpha}{2\pi}
\Biggl\{
\left[
\frac{x'}{x}[h_L(x) - K(x)] - \widetilde M (x)
\right]
\Theta_{11}^0 (x', x' - x)\nonumber\\
&+&
\left[
\frac{x}{x'}[h_L(x') - K(x')] - \widetilde M (x')
\right]
\Theta_{11}^0 (x, x - x')
\nonumber\\
&+&\int d\beta'
\Biggl(
\widetilde D(x, \beta')\frac{(\beta' - x + x')}{(x - x')}
\Theta_{111}^0 (x', x' - x , x' - x + \beta')\nonumber\\
&+&\frac{x'}{x' - \beta'}
[\widetilde D (x - x' + \beta', \beta') -\widetilde D (x, x')]
\Theta_{11}^0 (x' , x' - \beta')
\Biggr) \nonumber\\
&+&\int d\beta
\Biggl(
\widetilde D(\beta , x')\frac{(\beta - x' + x)}{(x' - x)}
\Theta_{111}^0 (x, x - x' , x - x' + \beta)\nonumber\\
&+&\frac{x}{x - \beta}
[\widetilde D (\beta , x' - x + \beta) -\widetilde D (x, x')]
\Theta_{11}^0 (x , x - \beta)
\Biggr) - \frac{3}{2}\widetilde D(x,x')
\Biggr\}.
\end{eqnarray}
The anomalous dimensions calculated from the evolution equation
for the distribution $h_1(x)$ coincide (up to the colour group
factor $C_F$) with the result of Ref. \cite{art90}.

By exploiting the relation provided by the equation of motion
and Lorentz invariance, we can easily verify that the RG equations
thus constructed are indeed correct and the renormalization
program can be reduced to the study of logarithmic divergences
of the three-parton ($Z(x,x')$, $\widetilde Z(x,x')$) and quark
mass ($M(x)$, $\widetilde M(x)$) correlators in perturbation theory.

\setcounter{equation}{0}

\section{Local anomalous dimensions}

In this appendix we pass from the evolution equations for
correlators to the equations for their moments and, in this
way, the anomalous-dimension matrix for local twist-3
operators.

We define the moments as follows:
\begin{eqnarray}
\label{mom-2-3}
F_n &=& \int dx x^n F(x)\hspace{0.5cm}\mbox{for any two-particle
correlator,}\nonumber\\
\label{defmom}
Z_n^l &=& \int dx dx' x^{n-l} x'^{l-1} Z (x,x').
\end{eqnarray}
In the language of operator product expansion these equalities
specify the expansion of non-local string operators in towers of
local ones, namely:
\begin{eqnarray}
{\cal Z}_n^l &=&
i^{n-1}(-1)^{l-1}
\frac{\partial^{l-1}}{\partial\mu^{l-1}}
\frac{\partial^{n-l}}{\partial\lambda^{n-l}}
\left.{\cal Z}(\lambda , \mu)\right|_{\lambda=\mu=0} \nonumber\\
&&\hspace{5cm}= \frac{1}{2} \bar \psi (0) (iD_+)^{l-1}
{\rm g} G_{+ \rho} (0) \sigma^\perp_{\rho +}
\!\left( \!
\begin{array}{c}
I \\
\gamma_5
\end{array}
\!\right)
(iD_+)^{n-l}\psi (0),\nonumber\\
{\cal M}_n &=&
i^n \frac{\partial^n}{\partial\lambda^n}\,
\left.{\cal M}(\lambda)\right|_{\lambda=0}
=\frac{m}{2}\bar \psi (0)
\gamma_+
\!\left( \!
\begin{array}{c}
I \\
\gamma_5
\end{array}
\!\right)
(iD_+ )^n \psi (0).
\end{eqnarray}
The inverse transformations to the non-local representation are
given by
\begin{eqnarray}
{\cal Z}(\lambda,\mu)
=\sum_{n=0,\,m=0}^{\infty} (-i)^{n+m}
(-1)^m\frac{\mu^m}{m!}\frac{\lambda^n}{n!}\,
{\cal Z}_{n+m+1}^{m+1},\qquad
{\cal M}(\lambda)
=\sum_{n=0}^{\infty} (-i)^{n} {\lambda^n\over n!}\,{\cal M}_n.
\end{eqnarray}

Now it is a simple task to derive the algebraic equations
for the mixing of local operators under the change of the
renormalization scale from the evolution equations
(\ref{eveqz})--(\ref{mass-NL}). They are
\begin{eqnarray}
\dot M_n &=& \frac{\alpha}{2\pi}
{_{\scriptscriptstyle MM}\gamma}^n M_n,\\
\dot Z^l_n &=& \frac{\alpha}{2\pi}
\left\{
\left[ {_{\scriptscriptstyle ZM}\gamma}^n_{n-l+1}
\pm {_{\scriptscriptstyle ZM}\gamma}^n_l \right] M_n
+\sum_{k=1}^{n}{_{\scriptscriptstyle ZZ}\gamma}^n_{lk} Z_n^k
\right\},
\end{eqnarray}
where the anomalous dimensions are given by the expressions
\begin{eqnarray}
{_{\scriptscriptstyle MM}\gamma}^n
&=& - C_F \left( S_n + S_{n+2} \right), \\
{_{\scriptscriptstyle ZM}\gamma}^n_l
&=& \frac{2C_F}{l(l+1)(l+2)}, \\
\label{anomdimZ}
{_{\scriptscriptstyle ZZ}\gamma}^n_{lk}
&=& \frac{3}{4}C_F \delta (l - k)
+\frac{C_A}{2}
\left\{
\theta (l-k-1)\frac{(k+1)(k+2)}{(l-k)(l+1)(l+2)}
-\delta (l-k)\left[ S_{k-1} + S_{k+2} \right]
\right\}\nonumber\\
&+& \left( C_F - \frac{C_A}{2} \right)
\biggl\{
\theta (l-k-1)
\left[
\frac{2(-1)^k C_l^k}{l(l+1)(l+2)}
+\frac{(-1)^{l-k}}{(l-k)}\frac{C_n^{k-1}}{C_n^{l-1}}
\right] \nonumber\\
&+& \delta (l-k)\left[
\frac{2(-1)^k}{k(k+1)(k+2)} -S_k
\right]
\biggr\}
+ {k \rightarrow n-k+1 \choose l \rightarrow n-l+1}.
\end{eqnarray}
Here we have used the following step functions
\begin{equation}
\label{Kron}
\theta (i-j)
= \left\{
\begin{array}{c}
1, \ i \geq j \\
0, \ i<j
\end{array}
\right. ,
\hspace{1cm}
\delta (i-j)
= \left\{
\begin{array}{c}
1, \ i = j \\
0, \ i \neq j
\end{array}
\right. ,
\end{equation}
as well as the convention
$ S_n = \sum_{k=1}^{n} \frac{1}{k}$ and the binomial coefficients
$C_n^m = \frac{n!}{m!(n-m)!}$.
The plus and minus signs in this equation correspond to the
functions $e$ and $h_L$, respectively. These analytical
expressions coincide with the result of Ref. \cite{koi95}.

\newpage

\vspace{2cm}

\unitlength1mm

\vspace{2cm}

\end{document}